\documentclass[10pt,journal,compsoc]{IEEEtran}
\usepackage{amsmath,amssymb,amsfonts}
\usepackage{pifont}
\usepackage[linesnumbered,ruled]{algorithm2e}
\usepackage{algorithmic}

\usepackage{textcomp}
\usepackage{xcolor}
\usepackage{makecell}
\usepackage{threeparttable}
\usepackage{graphicx}
\usepackage{multirow}
 \usepackage{booktabs} 
\usepackage{caption}
\usepackage{fixltx2e}
\usepackage{amsmath}
\usepackage{parskip}
\usepackage{lipsum}
\usepackage{wrapfig}

\usepackage{url}
\newcommand{\cmark}{\ding{51}} 
\newcommand{\xmark}{\ding{55}} 

\ifCLASSOPTIONcompsoc
  \usepackage[nocompress]{cite}
\else
  \usepackage{cite}
\fi
\ifCLASSINFOpdf
\else
\fi
\begin{document}
%
\title{SendingNetwork: Advancing the Future of Decentralized Messaging Networks}

\author{
    \IEEEauthorblockN{Mason Yeung} \\
    \IEEEauthorblockA{
        \textit{Sending Labs, United States}\\
    }
    \IEEEcompsocitemizethanks{\IEEEcompsocthanksitem This version was updated in January, 2024.}
}

%
%

\markboth{}%
{}
%



\IEEEtitleabstractindextext{%
\begin{abstract}
In the evolving landscape of Internet technologies, where decentralized systems, especially blockchain-based computation and storage like Ethereum Virtual Machine (EVM), Arweave, and IPFS, are gaining prominence, there remains a stark absence of a holistic decentralized communication framework. This gap underlines the pressing necessity for a protocol that not only enables seamless cross-platform messaging but also allows direct messaging to wallet addresses, fostering interoperability and privacy across diverse platforms. SendingNetwork addresses this need by creating a reliable and secure decentralized communication network, targeting essential challenges like privacy protection, scalability, efficiency, and composability. Central to our approach is the incorporation of edge computing to form an adaptive relay network with the modular libp2p library. We introduce a dynamic group chat encryption mechanism based on the \textit{Double Ratchet} algorithm for secure communication and propose a Delegation scheme for efficient message processing in large group chats, enhancing both resilience and scalability. Our theoretical analyses affirm the Delegation scheme's superior performance. To bolster system stability and encourage node participation, we integrate two innovative consensus mechanisms: "Proof of Relay" for validating message relay workload based on the novel KZG commitment, and "Proof of Availability" for ensuring network consistency and managing incentives through Verkle trees. Our whitepaper details the network's key components and architecture, concluding with a roadmap and a preview of future enhancements to SendingNetwork.
\end{abstract}

}

\maketitle

\IEEEdisplaynontitleabstractindextext

%
\IEEEpeerreviewmaketitle

\IEEEraisesectionheading{\section{Introduction}\label{sec:introduction}}
\IEEEPARstart{T}{he} advent of Internet technologies heralds a transformative era in digital communication and data management, characterized by a paradigm shift towards decentralization \cite{murray2023promise}. Central to this evolution is the integration of blockchain technology, which has revolutionized not only financial transactions but also the broader domain of computational tasks and data storage. Pioneering platforms like the Ethereum Virtual Machine (EVM) \cite{hildenbrandt2018kevm} have demonstrated the efficacy of blockchain in computational processes, while distributed storage solutions such as Arweave~\cite{williams2019arweave} and the InterPlanetary File System (IPFS) \cite{psaras2020interplanetary} have redefined data storage paradigms. 

Despite these remarkable advancements in decentralized computation and storage, the digital ecosystem still faces a notable void in one critical area: a comprehensive framework for decentralized communication. Communication is the backbone of the digital world, and its centralization poses significant risks and limitations, particularly in terms of privacy, security, and data sovereignty. This gap in the decentralized landscape is not just a technological challenge, but also a barrier to the broader adoption and maturation of blockchain technologies. Recognizing the urgency of this need, we present SendingNetwork, a decentralized communication network that is both robust and secure. The primary function of SendingNetwork is to facilitate the secure exchange of information in a public decentralized messaging network, ensuring the self-sustainability of the relay network and incentivizing nodes for timely message relay. In this system, messages are generated and transmitted off-chain, whereas the economic incentives and proof mechanisms are anchored and processed on the blockchain. Sendingnetwork enhances communication between blockchain accounts both within and across applications, efficiently tackling challenges in current messaging solutions:

\begin{itemize}
  \item \textbf{Identity Control:} Conflicts with social media entities often lead to account bans or deactivations, undermining online identity control \cite{jhaver2021evaluating}.

 \item \textbf{Identity Fragmentation:} Users' reliance on specific platforms curtails their mobility across different services and necessitates the maintenance of multiple platform accounts, leading to fragmented and complicated online identity management \cite{yang2020zero}.
  
  \item \textbf{Data Ownership and Privacy:} Modern applications risk exposing message histories to providers, despite encryption, enabling data mining and third-party exposure \cite{isaak2018user}. Platform shutdowns can result in significant data loss \cite{russell2013mining}. Additionally, the absence of end-to-end encryption in traditional messaging apps poses a risk to user data privacy and security.
  
  \item \textbf{Transaction Security:} The gap between social platforms and the crypto sphere, as seen in Discord and Whatsapp, impacts the verification of counterparties in transactions, raising security concerns.
\end{itemize}

\subsection{Related Works and Motivation}
Existing decentralized solutions still struggle to strike the optimal balance among scalability, privacy protection, and decentralization. In the realm of decentralized communication, a range of solutions has emerged, each attempting to address bottlenecks and concerns within communication networks. We primarily observe the following mainstream solutions.

\subsubsection{Peer-to-Peer Messaging}
Initially, messaging systems relied on a \textit{client/server} architecture, where a central server managed tasks such as account management, messaging services, storage, and more. In contrast, Peer-to-Peer networks distribute resources, with each node acting as both provider and consumer, thus enhancing scalability and reducing dependency on a single messaging server. This shift enables local data storage, granting users more control over their data.

Early decentralized protocols like Internet Relay Chat (IRC) \cite{shihab2009use} offered simple, text-based communication with low bandwidth requirements. The XMTP protocol represents another implementation in this field, utilizing wallet addresses for user identification and highlighting the concept of unified identity ownership. At present, its design is more focused on individual chats rather than on encrypting group conversations.

P2P systems face challenges such as network scaling issues \cite{scalability_p2p_systems} and lack of end-to-end encryption, posing security and privacy risks. Compared to modern systems, P2P networks often lack support for advanced features like multiple devices syncing and multimedia content, and the integration of encryption further complicates their functionality. These limitations have impacted their widespread adoption.

\subsubsection{Federated Messaging}
Federated nodes collectively create a network by hosting segments and acting as bootstrap nodes for new client connections. This approach, exemplified by Matrix \cite{jacob2019glimpse} and XMPP \cite{schuster2014global}, tackles scalability but inherits privacy concerns from centralized systems, as node owners can access private data and control server content. 

The model's reliance on federated nodes for peer discovery and key exchange \cite{peer_to_peer_social_networks} presents a challenge in striking the right balance between decentralization and scalability. In addition, messaging clients rely extensively on servers for storing account details and communication history. This dependency can result in service disruptions during server downtime, potentially leading to the loss of user identities and complete communication histories if the chosen server experiences an outage.

\subsubsection{Blockchain-backed Messaging}
Blockchain introduces the concept of wallet addresses, which function as unique, universal cross-platform identifiers, fundamentally ensuring that users maintain complete ownership and control over their digital identities. This shift towards wallet-based identity systems represents an advancement in the pursuit of authentic user identity ownership, aligning with the principles of decentralization and user autonomy.

As the emergence of blockchain-based solutions leveraging public ledgers for message storage becomes more prevalent, they encounter notable challenges, particularly regarding latency issues and privacy issues. Additionally, the inherent immutability of blockchain means that once messages are stored, they are permanently recorded. This characteristic of blockchain, while beneficial for transparency and security, poses a dilemma for private messaging, where the permanent storage of messages may not be desirable or appropriate. These challenges are a direct consequence of the inherent limitations that are characteristic of blockchain. This aspect of blockchain functionality makes it less optimal for applications requiring large-scale, real-time communication. The latency issue fundamentally restricts the efficacy of these blockchain solutions in scenarios where immediate data transmission and high-speed interactions are critical, thus limiting their applicability in certain real-time communication contexts \cite{liu2020blockchain}.



\subsubsection{Motivation}
As we examine the history of decentralized chat, SendingNetwork acknowledges the valuable lessons learned from these approaches. SendingNetwork's rationale lies in delivering a decentralized chat solution that excels in scalability, security, reliability, and usability. It aims to build upon the strengths while addressing the limitations. Our system combines the Matrix Protocol's strengths with original advancements across account systems, instant messaging, and data storage, resulting in an efficient decentralized chat solution.

\begin{table*}[htbp]
\scriptsize
    \centering
    \begin{threeparttable}
    \begin{tabular}{llccccccccc}
        \toprule
        
        & \multicolumn{4}{c}{Decentralization\textsuperscript{a}}  &  \multicolumn{3}{c}{Privacy\textsuperscript{b}}  &  \multicolumn{2}{c}{Scalability\textsuperscript{c}} \\
        
        \cmidrule(lr){2-5} \cmidrule(lr){6-8} \cmidrule(lr){9-10}
        
        Network Scheme  & 
        	IR & 
	IMSR & 
	IM & 
	CAS &
	DMEE & 
	GMEE &
	ZPDS & 
	VCPM &
	BSM \\
	        
        \midrule
        
        Edge Network & \cmark & \cmark & \cmark & \cmark & \cmark & \cmark & \cmark & \cmark & \cmark \\
        
        P2P & \cmark & \cmark & \xmark & \cmark & \cmark & \xmark & \xmark & \xmark & \xmark \\
        
        Federated  & \xmark & \xmark & \xmark & \xmark  & \cmark & \cmark & \xmark & \cmark & \cmark \\
     
        \bottomrule
    \end{tabular}
    
     \begin{tablenotes} [para,flushleft]
        \scriptsize
        \item \textsuperscript{a} \textit{IR} Identity Resilience, \textit{IMSR} IM Service Resilience, \textit{IM} Incentives Mechanism, \textit{CAS} Crypto Account Support
        
        \item \textsuperscript{b} \textit{DMEE} Direct Messge End-to-end Encryption, \textit{GMEE} Group Message End-to-end Encryption, \textit{ZPDS} Zero Private Data Storage
        
        \item \textsuperscript{c} \textit{HRE} High Relay Efficiency, \textit{BSM}  Broadcast Storm Mitigation
      \end{tablenotes}
    \end{threeparttable}
    
    \caption{Decentralized Messaging Solution Status} 

\end{table*}

\subsection{Design Principles}
The architecture of SendingNetwork is strategically developed to provide a stable and decentralized messaging platform. At its core, the design principles prioritize scalability, privacy, and composability within a decentralized context. The fundamental
design principles are defined as follows.
\begin{itemize}
\item \textbf{Permissionless Engagement:} Messaging clients no longer rely on trusting any third party or being under the control of any centralized organization; anyone can participate as a relay node within the network. The blockchain's \textit{Proof of Availability} and \textit{Proof of Relay} mechanisms ensure the network's complete decentralization.
\item \textbf{Self-Sustainable Network:}
The relay network employs multiple nodes for each messaging client, mitigating the risk of a single point of failure. Consequently, even if certain nodes become inoperative, client access and data integrity remain unaffected due to certain node redundancy and backup.

To bolster network resilience, nodes are dynamically allocated, guaranteeing continuous access for messaging clients, and maintaining network functionality and client service even when some nodes are temporarily offline. Furthermore, the \textit{Proof of Relay} mechanism plays a crucial role in dynamically adjusting node incentives. In situations where there is a shortage of node resources, the relay rewards are increased to attract additional nodes, ensuring the sustained efficiency of the relay network.

\item \textbf{Intrinsic Data Privacy:}
Data privacy within the network is rigorously maintained through a series of strategic measures. Foremost is the implementation of end-to-end encryption, utilizing the \textit{Double Ratchet} algorithm for efficient key exchange, particularly crucial in the context of large group chats. This ensures that message content remains accessible solely to the conversation participants, thereby securing privacy and blocking unauthorized access.

Additionally, the network employs secure data caching for offline clients. Relay nodes temporarily cache messages, all of which are end-to-end encrypted to maintain security and restrict access from unauthorized entities.

A key aspect of maintaining data integrity is the verifiability of the information transmitted across the network. Each message is signed with the client’s private key, and upon receipt, these messages undergo rigorous verification to confirm their authenticity. This robust verification process is designed to promptly detect and reject any data that has been tampered with, thus ensuring the reliability and authenticity of the information within the network.

\item \textbf{Scalable Communication Infrastructure:}
Utilizing \textit{libp2p} for a multicast strategy effectively reduces relay instance frequency, addressing network congestion issues. This approach hinges on a \textit{Pub/Sub} mechanism for relaying messages, eschewing the need for broad network broadcasting. 

For large-scale public communities, the Delegation scheme reduces network traffic pressure and ensures uninterrupted user service during node failures. It enhances efficiency by serving multiple messaging clients publicly, effectively reducing network traffic. It achieves this by transmitting messages directly to connected clients, thus bypassing the necessity of relaying through the broader network.

This configuration enables comprehensive end-to-end encryption for extensive group communications. For private messaging, users access the relay network directly via local client nodes, bypassing the need for third-party involvement. Conversely, Delegation client nodes facilitate access for large-scale group chats, achieving a balance between cost-efficiency and performance in large group settings.

\item \textbf{Data Minimization: }
Data minimization in a relay network adheres to a strategy that restricts data collection and storage to the essentials required for network functionality. This approach serves a dual purpose: it reduces the load on network resources, making it leaner and faster, and it significantly lessens the risk of data breaches, thereby enhancing user privacy.

In the relay network, decentralized room state synchronization, similar to Git \cite{arndt2019decentralized}, removes the need for node-stored chat histories. This allows client devices to independently store data, enabling ephemeral message transmission within the network. Messages are not stored permanently but exist transiently, reducing dependency on external data storage. For offline clients, edge nodes temporarily cache messages, which are deleted upon client reconnection.
\end{itemize}

\subsection{Contributions}
The main contributions are summarized as follows.
\begin{itemize}
    \item First of all, we propose a groundbreaking, decentralized three-tier (access-relay-consensus) messaging architecture, addressing the core limitations of existing solutions. This architecture leverages edge computing to develop an adaptive, dynamic message relay network, offering scalability and efficiency through the underlying libp2p stack and a temporary caching mechanism. 
    \item Second, our framework features an advanced dynamic group chat encryption method, grounded in the \textit{Double Ratchet} algorithm. This mechanism ensures robust end-to-end data privacy and integrity, safeguarding user communications against unauthorized access and tampering.
    \item Third, we present an innovative Delegation scheme, specifically designed for processing messages in large group chats, offering resilience protection while minimizing network traffic. Theoretical analysis demonstrates the reduced communication complexity of our scheme compared to traditional broadcast messaging, suggesting a significant potential for further scalability enhancement.
    \item Finally, we establish two customized consensus mechanisms, called \textit{Proof of Relay} and \textit{Proof of Availability}, to verify the relay operations, preserve global state consistency, and determine the order of bill settlement.
\end{itemize}
\subsection{Organization}
The rest of the paper is organized as follows. In Section~\ref{sec:bk}, the fundamental knowledge utilized in SendingNetwork is explained. In Section~\ref{sec:so} and Section~\ref{sec:cc}, the system model and concrete constructions are described, respectively. In Section~\ref{sec:cm}, the designed consensus mechanisms are detailed. Finally, we outline the future work and conclude the main features of SendingNetwork in Section~\ref{sec:fw} and Section~\ref{sec:cl}, respectively.

\section{Background Knowledge}
\label{sec:bk}
This section elucidates the definitions of three fundamental technical components of SendingNetwork, aimed at simplifying the reader's comprehension.
\subsection{Bilinear Pairing}
Let $\mathbb{GF}(p)$ denote a finite field, where $p$ is a large prime number. We consider an elliptic curve $E_p(a,b)$ over $\mathbb{GF}(p)$, which can be defined as the set of ordered pairs $(x,y) \in \mathbb{GF}(p) \times \mathbb{GF}(p)$ that fulfill the condition $y^2 \equiv x^3 + ax + b \;(\text{mod}\; p)$, given that $a, b \in \mathbb{GF}(p)$ and $4a^3 + 27b^2 \not\equiv 0\; (\text{mod}\; p)$. This leads to the construction of an additive cyclic group, denoted $\mathbb{G}_1$, and a multiplicative cyclic group, $\mathbb{G}_2$, both of which have the prime order $p$. These groups are constituted of all points residing on the elliptic curve, complemented by the point at infinity. A random generator of $\mathbb{G}_1$ is designated as $P$. We define a bilinear map $e: \mathbb{G}_1 \times \mathbb{G}_1 \rightarrow \mathbb{G}_2$ (known as a type-1 pairing) that adheres to the following three properties \cite{meffert2009bilinear}:
\begin{itemize}
    \item \textbf{Bilinearity:} For any $X, Y \in \mathbb{G}_1$ and any $a, b \in \mathbb{Z}_p^\ast$, it holds that $e(aX, bY) = e(X, Y)^{ab}$.

\item \textbf{Non-degeneracy:} If we designate $1_{\mathbb{G}_2}$ as the identity element of $\mathbb{G}_2$, it should be ensured that $e(X, Y) \neq 1_{\mathbb{G}_2}$ for any $X, Y \in \mathbb{G}_1$.

\item \textbf{Computability:} For any $X, Y \in \mathbb{G}_1$, the value $e(X, Y)$ can be efficiently computed.
\end{itemize}

Two fundamental problems play a crucial role in ensuring the robustness of our protocol \cite{galbraith2016recent}. These problems, known as the Elliptic Curve Discrete Logarithm Problem (ECDLP) and the Elliptic Curve Computational Diffie-Hellman Problem (ECCDHP), are outlined below:

\begin{itemize}
    \item \textbf{ECDLP:} Given an elliptic curve $E_p(a,b)$ over a finite field $\mathbb{GF}(p)$, and two points $P, Q \in \mathbb{G}_1$, the ECDLP is to find an integer $x$ such that $Q = xP$ if one exists. It is conjectured that no efficient algorithm can solve this problem, making it the foundation of the security in elliptic curve cryptography. The intractability of ECDLP ensures that, even if an attacker knows $P$ and $Q$, they cannot feasibly compute the integer $x$.

\item \textbf{ECCDHP:} For an elliptic curve $E_p(a,b)$ over a finite field $\mathbb{GF}(p)$, and given $P, aP, bP \in \mathbb{G}_1$, the ECCDHP asks for the computation of the point $abP$. In the realm of cryptographic systems, ECCDHP is considered computationally hard, assuming the hardness of the ECDLP. In simpler terms, if one can solve ECCDHP efficiently, then they can also solve ECDLP efficiently, which is conjectured to be infeasible.
\end{itemize}
\subsection{Double Ratchet Algorithm}
The \textit{Double Ratchet} mechanism \cite{perrin2016double, alwen2019double}, a pivotal concept in the field of cryptographic communications, epitomizes an advanced methodology for ensuring the privacy and security of messaging protocols. It operates on the foundational principle of combining two distinct ratcheting processes: the Diffie-Hellman ratchet and the Symmetric-Key ratchet. The Diffie-Hellman ratchet, named after its creators, facilitates the establishment of new shared secrets between communicating parties at regular intervals, thereby imbuing the communication channel with forward secrecy. Each message transmission triggers the generation of new Diffie-Hellman key pairs, ensuring that the compromise of current keys does not jeopardize the confidentiality of past communications. Complementing this, the Symmetric-Key ratchet is responsible for updating the encryption key for each message. By continuously evolving the key with each exchanged message, it provides future secrecy, ensuring that the compromise of a single key does not lead to the unraveling of subsequent message contents. The amalgamation of these two mechanisms in the \textit{Double Ratchet} algorithm results in a robust cryptographic system that dynamically updates encryption keys, thereby significantly enhancing the security of ongoing communication streams against potential threats and vulnerabilities.

\subsection{libp2p Library}
The libp2p library, an integral component of modern decentralized network architectures, is a multifaceted and modular networking stack designed for peer-to-peer (P2P) applications \cite{guidi2021libp2p}. As a cornerstone of the InterPlanetary File System (IPFS) \cite{psaras2020interplanetary} and other decentralized systems, libp2p offers a comprehensive suite of protocols and tools for building complex, scalable, and interoperable networked applications. The core features can be summarized as follows.
\begin{itemize}
    \item Modularity and Extensibility: At its core, libp2p is distinguished by its modular design, allowing developers to choose and integrate only the components necessary for their specific use case. This modularity spans various network functions such as transport, peer discovery, encryption, stream multiplexing, and protocol negotiation.
\item Transport Agnosticism: libp2p abstracts the transport layer, supporting multiple transport protocols like TCP, UDP, WebSockets, and others. This flexibility enables applications to run seamlessly over different network configurations and conditions.
\item Peer Discovery and Routing: libp2p incorporates sophisticated peer discovery mechanisms, enabling nodes to efficiently find and connect with each other in a distributed network environment. It also provides advanced routing capabilities, essential for efficient data propagation in decentralized systems.
\item Encryption and Security: Security is a fundamental aspect of libp2p, featuring built-in support for encrypted connections. This ensures secure communication channels between peers, safeguarding data integrity and confidentiality.
\item NAT Traversal: Addressing common challenges in P2P networking, libp2p includes mechanisms for NAT traversal, facilitating direct connections between peers in different network environments.
\item Stream Multiplexing: libp2p allows multiple virtual streams to be multiplexed over a single physical connection, enhancing network efficiency and reducing resource overhead.
\item Protocol Agnosticism and Upgradeability: The library supports a variety of application-level protocols and allows for seamless protocol upgrades, catering to evolving network requirements and standards.
\item Interoperability and Cross-Platform Support: libp2p is designed for cross-platform compatibility, ensuring interoperable communication across different implementations and programming languages.
\end{itemize}

 Its modular architecture, combined with a comprehensive set of networking features, positions libp2p as a pivotal tool for building the next generation of decentralized communication.

\section{System Overview}
\label{sec:so}
\begin{figure*}[h]
    \centering
    \includegraphics[width=0.8\textwidth]{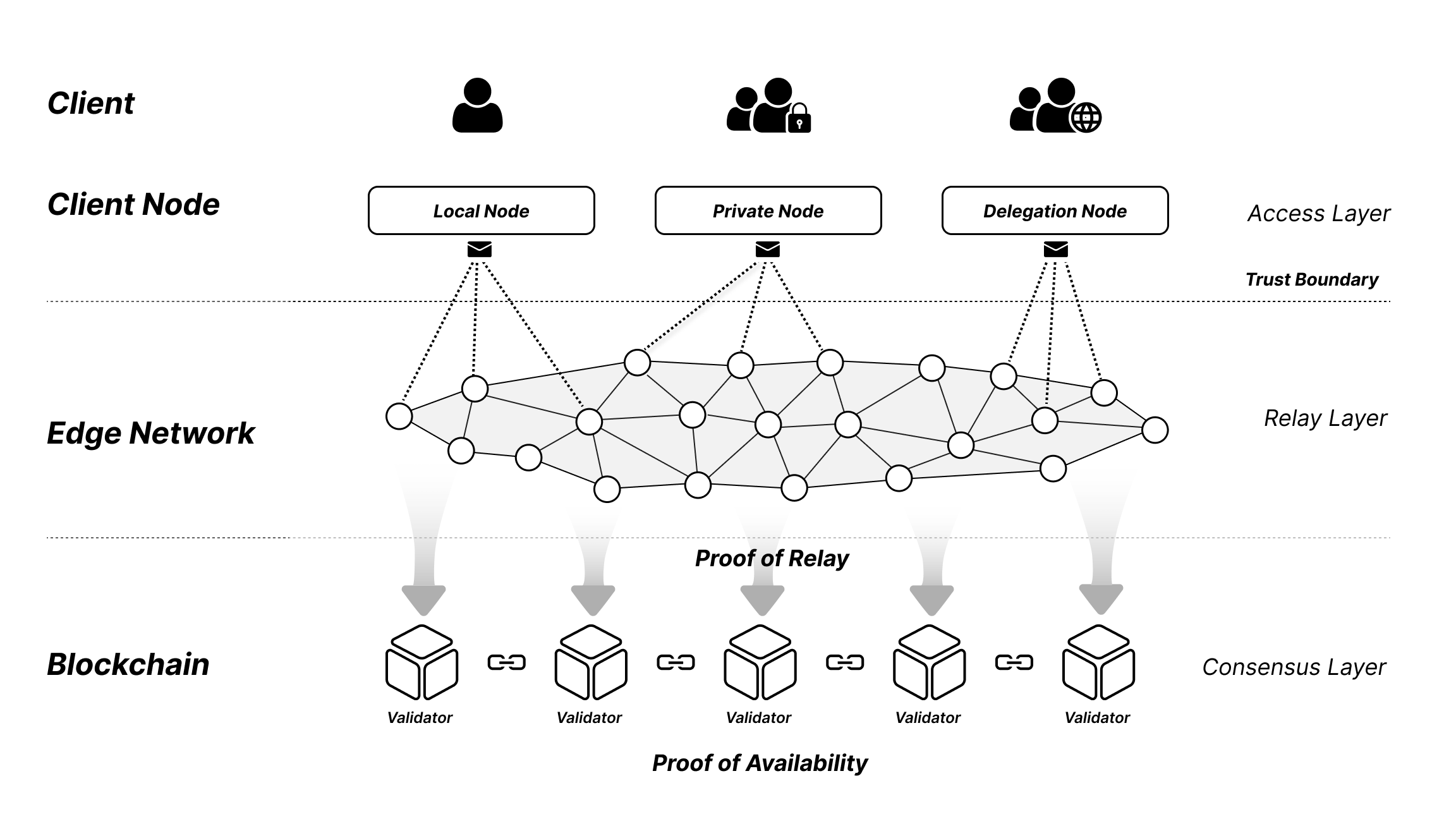}
    \caption{Protocol Architecture}
    \label{fig:protocol_architecture}
\end{figure*}

The proposed real-time communication protocol merges blockchain, P2P networking, end-to-end encryption, delegation, and related technologies to establish a self-sustaining, secure, and highly efficient system. This protocol is structured into three distinct layers, visually represented in Figure \ref{fig:protocol_architecture}:

\begin{enumerate}
  \item Access Layer: Engages the user client, facilitating P2P edge network entry via the client node.
  \item Relay Layer: Encompasses the edge network, comprising edge nodes delivering instant messaging services without user data retention.
  \item Consensus Layer: Validators utilize \textit{Proof of Availability} to establish global economic consensus while messaging participants employ \textit{Proof of Relay} for settling messaging bills.
\end{enumerate}

\subsection{Key Components}
Four pivotal roles in P2P communication emerge: client, client node, edge node, and validator.

\subsubsection{Client}
Functioning as the user terminal, clients represent applications interfacing with client nodes via client/server communication, typically integrating the SDK. They incorporate end-to-end encryption employing the \textit{Double Ratchet} algorithm, responsible for encryption session maintenance and key exchange.

Operating on user-trusted and controlled devices like mobile phones, tablets, or PCs, the client securely stores fundamental user data. Access to the P2P relay network occurs through client nodes like local P2P nodes or remote Delegation nodes. These nodes handle message processing, transmission, and reception, akin to traditional IM servers. Both the client SDK and client nodes hold user trust.

Users may access the network via local or Delegation nodes. For those hesitant about relying on third-party Delegation nodes, direct P2P network access is attainable through local nodes or personalized, privately hosted Delegation nodes.

\subsubsection{Client Node}
The network architecture consists of client nodes, encompassing local nodes, Delegation nodes, and private Nodes, acting as intermediaries linking edge nodes within the relay network to clients. These nodes facilitate client-server model-based communication. Messages received by client nodes are sent to edge nodes for transmission to the recipient's client nodes, which then forward them to the intended recipients.

In contrast to edge nodes, which operate without the necessity of trust, client nodes are entrusted by clients. As edge nodes do not offer persistent storage for clients, all client communication history is encrypted and securely stored in client nodes. There exists no fixed binding between clients and client nodes, allowing users to seamlessly transition to a new client node while maintaining uninterrupted messaging capabilities. Users can also back up their message history and restore it on another client node, ensuring uninterrupted communication.

Client Nodes fall into three categories based on their relative positioning to clients, the client range they serve, and their service offerings.

\begin{itemize}
  \item \textit{Local Nodes} cater exclusively to local clients, operating on a client's device and storing all communication history locally. These nodes are preferred by users with stringent privacy requirements due to the local storage of communication data.
  
  \item \textit{Delegation Nodes} operate remotely and extend their services to the wider public, catering to all public clients. Clients have the liberty to connect to a delegation node instead of hosting a local node, considering factors such as performance, battery consumption, or storage limitations. These nodes are similar to Ethereum's RPC nodes. 
  
  \item \textit{Private Nodes} are specialized delegation nodes with access control, servicing a dedicated set of clients. Typically hosted by developers integrating SendingNetwork for messaging services, these nodes are exclusively utilized to serve specific user bases, facilitating customized messaging solutions.
\end{itemize}

\subsubsection{Edge Node}
Edge Nodes hold pivotal roles in message relay and data caching. Their function includes message relay rewards, akin to being the network's ``miners''.

The core resources of edge nodes encompass bandwidth and caching, facilitating the caching of encrypted messages for numerous offline clients. They establish high-speed communication channels with online clients, enabling not only message relay but also seamless streaming media transmission. Messages delivered to edge nodes are encrypted and signed, ensuring the confidentiality and integrity of user data. Offline data cached on edge nodes is promptly deleted upon retrieval by offline clients upon reconnection.

These nodes handle text, voice, and video messaging and can extend their capabilities to process text-based smart contract commands. This allows users to execute transactions while they are engaged in messaging activities.

\begin{figure*}[ht]
    \centering
    \includegraphics[width=0.8\textwidth]{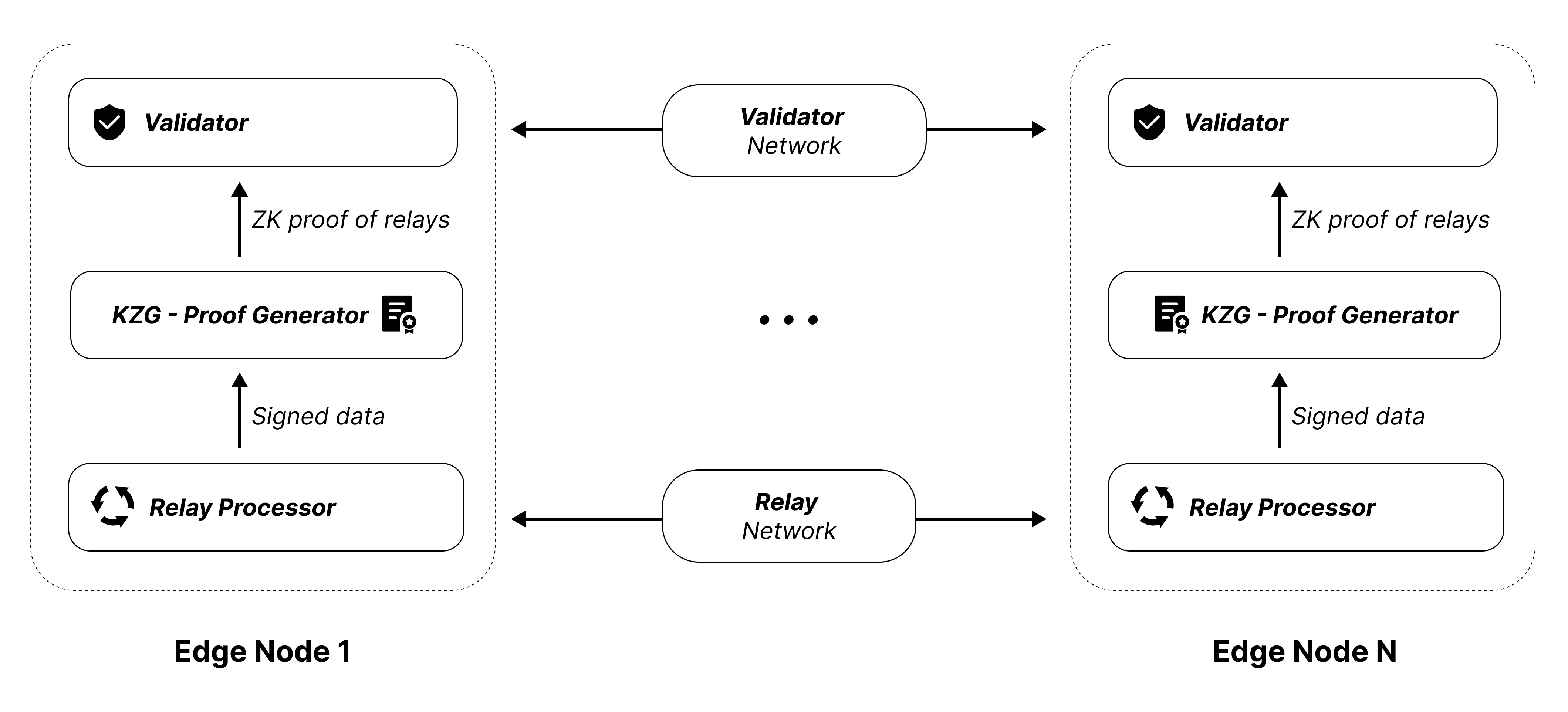}
    \caption{Edge Node Framework}
    \label{edge_node_component}
\end{figure*}

Within the network, edge nodes adhere to the P2P protocol, enabling server-to-server communication. For an intricate breakdown of an edge node's components, please refer to Figure~\ref{edge_node_component}.

\subsubsection{Validator}
Validator plays a pivotal role in overseeing network operations and ensuring security within SendingNetwork. Utilizing the \textit{Proof of Availability} consensus mechanism, the network selects validators from edge nodes. Once an edge node becomes a Validator, it assumes the responsibility of generating new blocks containing zk proofs derived from the \textit{Proof of Relay} workload proof of edge nodes. Simultaneously, validators monitor the recent online status of edge node peers, incentivizing them to consistently provide stable message relay services.

The resultant blockchain operates as a registrar for user registration data and coordinates edge node resources within the network. Its primary functions involve managing user Decentralized Identifiers (DID) information, organizing and incentivizing edge node resources, and motivating miners to contribute bandwidth, fostering the network's operation within a fully decentralized environment. 


\subsection{Building Blocks}
\subsubsection{Adaptive Edge Network}
We expand upon the P2P network by introducing a decentralized and security-protected edge computing architecture. Specifically, each client node is randomly allocated multiple edge nodes, tasked with message relay and temporary caching for offline clients. Consequently, rather than broadcasting messages across the network, the client node assigns the messages to its edge nodes, and edge nodes will route messages to the recipient's client nodes. This approach reduces the number of message relays, diminishing forwarding frequencies. This network model notably truncates peer discovery durations. All traffic, including messages and resource locating, operates within the network, resulting in a significant reduction in overall traffic. This mitigates the P2P broadcast storm issue, preventing message floods across the entire network.

Additionally, compared to the conventional P2P broadcasting mode, where message loss or increased latency occurs if the next hop isn't located, our approach circumvents these issues. In a P2P network, the likelihood of message loss can be affected by several elements, including node churn (frequency of nodes entering and exiting the network), network congestion, and node reliability. An improved network architecture would consider these aspects. Edge Nodes, driven by incentives to deliver consistent and dependable relay services, boast substantially higher uptime compared to standard client P2P nodes. This heightened reliability significantly influences the network's overall message delivery efficiency.

The network is a dynamic system where messages are continuously transmitted, and latency and re-transmissions are functions of time. We introduce integrals to represent the accumulation of latency and the impact of re-transmissions over time. We analyze the efficiency of the proposed edge-enabled relay network as follows.

$\blacktriangleright$ \textbf{Complexity Analysis:}

The utilized notations are defined in Table~\ref{tab:no0}. 

\begin{table}[htbp]
    \begin{center}
    \caption{Notation Descriptions}
    \label{tab:no0}
    \renewcommand{\arraystretch}{1.5}
    \begin{tabular}{|c|c|}
    \hline
    \textbf{Notation} & \textbf{Definition} \\
    \hline
    $M$ & Total number of messages to be sent \\
    \hline
    $N$  &  The total number of client nodes \\
     \hline
    \multirow{2}{*}{$k$}  & Fixed number of edge nodes assigned to \\ & receive messages for each client node \\
    \hline 
    $l(t)$ & The latency function \\
    \hline
   $p_{\text{loss}}(t)$ & The probability of message loss \\
    \hline
    $r(t)$ & The expected number of re-transmissions \\
    \hline
    \end{tabular}
    \end{center}
\end{table}


For the traditional broadcasting approach, the expected total time $E[T_{\text{broadcast}}]$ including variable latency and re-transmissions over the continuous transmission period $[0, T]$ would be:

\begin{equation}
\begin{aligned}
E[T_{\text{broadcast}}] & = \int_{0}^{T} M \cdot (N - 1) \cdot \\
& \Big(t_m + l(t)\Big) \cdot \Big(1 + r(t) \cdot p_{\text{loss}}(t)\Big) \, dt
\end{aligned}
\end{equation}

For our approach, the expected total time $E[T_{\text{edge\_network}}]$ including variable latency and re-transmissions over the same period would be:

\begin{equation}
\begin{aligned}
E[T_{\text{edge\_network}}] & = \int_{0}^{T} M \cdot k \cdot \Big(t_m + l(t)\Big) \cdot \\
& \Big(1 + r(t) \cdot p_{\text{loss}}(t)\Big) \, dt
\end{aligned}
\end{equation}

In these integrals, $T$ is the total time period over which messages are being sent. The term $(t_m + l(t))$ represents the instantaneous time to transmit a message at time $t$, including the instantaneous latency $l(t)$. The term $(1 + r(t) \cdot p_{\text{loss}}(t))$ accounts for the expected number of transmissions needed at time $t$ due to re-transmissions from message loss.

By comparing the integrals, we can argue that because $k$ is constant and much smaller than $(N - 1)$, and $l(t)$, $p_{\text{loss}}(t)$, and $r(t)$ are more favorable in the edge network scheme due to its efficient design, the integral for $E[T_{\text{edge\_network}}]$ would result in a lower value than that for $E[T_{\text{broadcast}}]$.

These integrals represent a continuous model of the network's behavior, accounting for variations in latency and message loss probability over time, which could be more reflective of a real-world dynamic network.

\subsubsection{Dynamic Group Encryption}
End-to-end encryption involves the encryption of message content, utilizing a symmetric encryption mechanism. The transmission of the encryption key employs asymmetric encryption methods.

Implemented through the \textit{Double Ratchet} algorithm, each member maintains a local keychain of other chat room participants. Messages undergo encryption using the client's key from the keychain, employing an incremental index. Subsequent decryption involves incrementing the keychain and using the index to verify the accurate symmetric encryption key within the keychain for message decryption.

Alterations in the group's composition, such as new user entry or existing member departure, prompt a complete re-exchange of encryption keys among all group members. This ensures the enabling of new users to encrypt/decrypt messages while restricting departed members' access to group communication.

Conventional end-to-end encryption methods encounter significant challenges in extensive group chats. For a group with \(N\) members, completing a key exchange requires approximately \(O(N^2)\) messages. In practical terms, a 500-member group necessitates at least 250,000 messages for key exchanges (500 * 500), potentially escalating to 1 million messages. Thus, employing end-to-end encryption in large groups is most effective when a minority actively engages in message transmission, while the majority passively receives messages.

Adaptive end-to-end encryption introduces a chat group threshold, beyond which pairwise key exchanges automatically shift to large group-specific encryption. Despite this transition, the encryption remains end-to-end. When the group surpasses this threshold and the encryption key update criteria are met, the initial sender generates a shared encryption key for the entire group, streamlining the process instead of each member exchanging keys individually.

\begin{figure*}[t]
    \centering
    \includegraphics[width=0.9\textwidth]{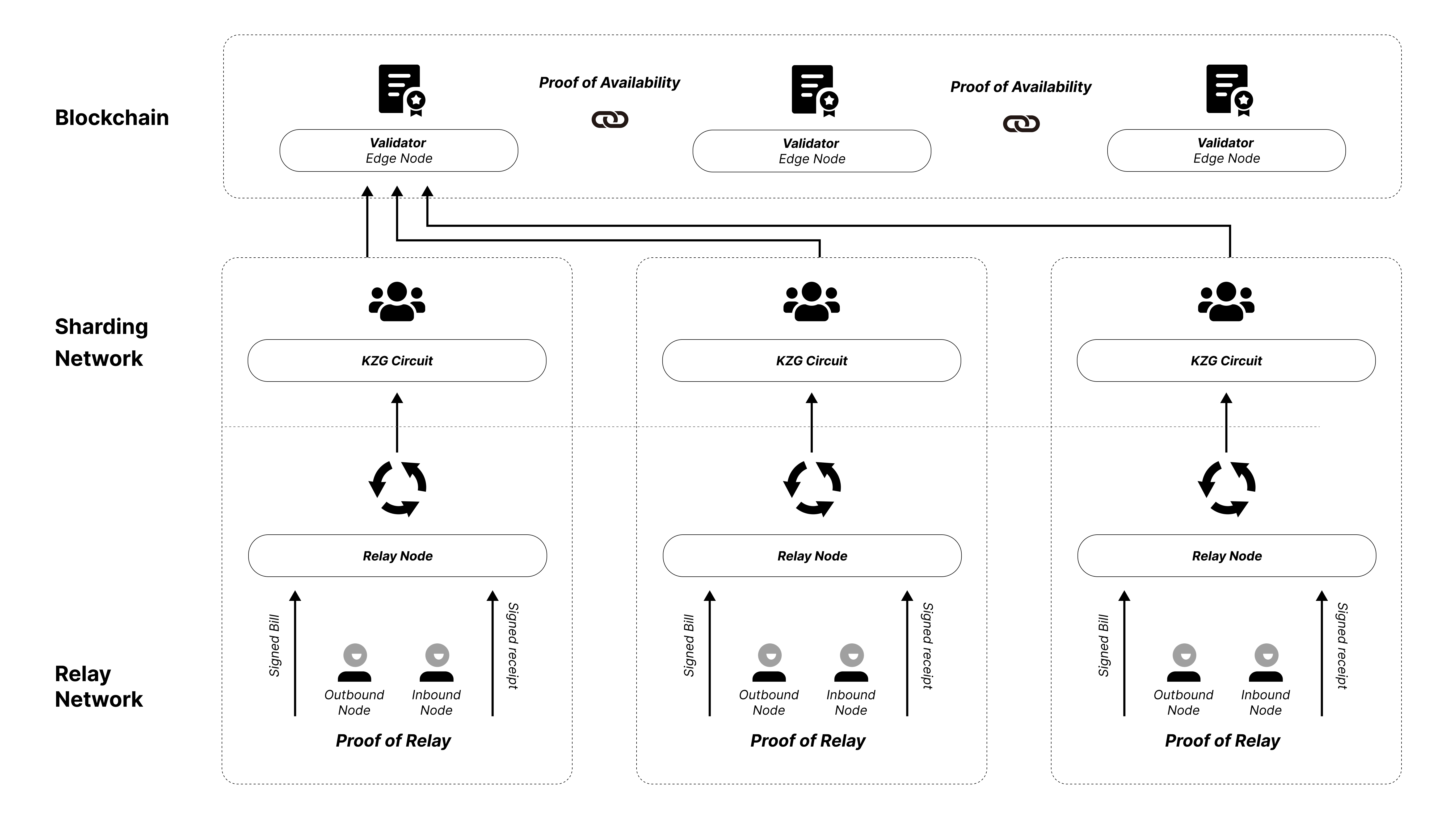}
    \caption{Blockchain Interactions}
    \label{fig:consensus}
\end{figure*}

\subsubsection{Delegation Scheme}
\label{delegation_scheme}
Delegation Nodes represent a distinct form of client node within the system. Unlike local nodes, these are publicly accessible, allowing client nodes to establish connections via node-specific URLs. The fundamental role of Delegation nodes lies in enhancing network scalability, particularly in supporting large-scale group conversations.

These nodes find utility primarily in instances where developers configure delegation nodes to provide services to their users. Users connect to these nodes to establish communication channels. In such scenarios, a significant proportion of users within a chat group, if not all, might be connected to a single delegation node. When a client connected to this node sends a message, the delegation node swiftly determines if the recipient is also connected to a client node associated with it, directly rerouting the message in that case.

Concurrently, upon receiving messages from clients, delegation nodes disseminate these messages to edge nodes subscribed to the corresponding chat room. The edge nodes, in turn, relay these messages to the client nodes of all recipients, inclusive of the delegation node itself. In the event of the delegation node receiving a message from an edge node, it automatically disregards the message since it has already forwarded it to the recipient's client. This operational model significantly reduces the number of messages within the network, notably diminishing the volume of ordinary messages, including those related to key exchanges, particularly when handling extensive encrypted group conversations.

Using multiple instances, Delegation nodes concurrently bolster the load capacity of the local node by implementing load-balancing configurations. Unlike permissionless edge nodes, delegation nodes are customizable, offering the possibility of configuring secure strategies to create permissioned access, thereby providing private services to specific clients.

$\blacktriangleright$ \textbf{Complexity Analysis:}

To illustrate the benefits of incorporating delegation nodes in group chats, we consider scenarios where certain group members might connect to the same Delegation node while others connect to distinct client nodes. The notation descriptions are defined in Table~\ref{tab:no}.

\begin{table}[htbp]
    \begin{center}
    \caption{Notation Descriptions}
    \label{tab:no}
    \renewcommand{\arraystretch}{1.5}
    \begin{tabular}{|c|c|}
    \hline
    \textbf{Notation} & \textbf{Definition} \\
    \hline
    $G$ & Total number of groups \\
    \hline
    $M_i$  &  Number of members in group $i$ \\
     \hline
    \multirow{2}{*}{$M_{i,d}$}  & Number of members in group $i$ \\
    & that are connected to Delegation node $d$ \\
    \hline 
    \multirow{2}{*}{$M_{i,o}$} & Number of members in group $i$ \\
    & that are connected to other nodes \\
    \hline
    \multirow{2}{*}{$D_i$} & Number of Delegation nodes that \\ 
    & members in group $i$ are connected to \\
    \hline
    \multirow{2}{*}{$T$} & Average number of messages sent by \\
    & each client node in a time interval \\
    \hline
    \end{tabular}
    \end{center}
\end{table}

Then, the total number of members in group $i$ is the sum of members connected to all delegation nodes plus those connected to other nodes:

\begin{equation}
M_i = \sum_{d=1}^{D_i} M_{i,d} + M_{i,o}
\end{equation}

Now, we can calculate the total number of message transmissions with the delegation scheme, taking into account the different connections of group members.

\textbf{With the Designed Delegation Scheme:}

When a client sends a message within a group, if the recipients are connected to the same delegation node, the message is transmitted only once by the delegation node to all those recipients. If the recipients are connected to different client nodes, each message must be transmitted individually to each of those nodes.

The total number of message transmissions with delegation nodes, $T_{\text{deleg}}$, can be expressed as:

\begin{equation}
T_{\text{deleg}} = \sum_{i=1}^{G} \left( T \cdot \sum_{d=1}^{D_i} (M_{i,d} + 1) + T \cdot M_{i,o} \right)
\end{equation}

This equation consists of two parts:

\begin{enumerate}
  \item $T \cdot \sum_{d=1}^{D_i} (M_{i,d} + 1)$: The number of transmissions within the delegation nodes for each group.
  \item $T \cdot M_{i,o}$: The number of transmissions to members connected to other nodes.
\end{enumerate}

\textbf{Improvement Factor:}

The improvement factor $I$ comparing the delegation model to the no-delegation model now becomes:

\begin{equation}
\begin{aligned}
I = & \frac{T_{\text{no\_deleg}}}{T_{\text{deleg}}} \\
 = & \frac{\sum_{i=1}^{G} M_i \cdot T \cdot (M_i - 1)}{\sum_{i=1}^{G} \left( T \cdot \sum_{d=1}^{D_i} (M_{i,d} + 1) + T \cdot M_{i,o} \right)}
\end{aligned}
\end{equation}

This improvement factor $I$ shows the reduction in the total number of message transmissions due to the delegation scheme. $T_{\text{no\_deleg}}$ exhibits a complexity relevant to $O(N^2)$, while $T_{\text{deleg}}$ is characterized by a complexity of $O(N)$. Consequently, $I$ typically yields a value greater than one, indicating that delegation reduces network traffic and improves scalability.

\subsubsection{Consensus Mechanisms}
SendingNetwork utilizes two consensus mechanisms to maintain the order of message fee settlement and guarantee the edge node availability. In this system, \textit{Proof of Availability} acts as the network's main spine, ensuring global consensus and the effectiveness of network governance; while \textit{Proof of Relay} handles specific network operations, improving network flexibility and response speed. This dual consensus mechanism design ensures that SendingNetwork is not only stable and reliable at the global level but also efficient and transparent in handling specific operations.

\begin{itemize}
    \item \textbf{Proof of Relay}: \textit{Proof of Relay} mechanism operates within a network for linear consensus, constituting a vital part of the network communication flow. Every message forwarding and processing adheres to fixed standards set by \textit{Proof of Relay}, ensuring transparency and verifiability for each operation. This process operates independently of any centralized trust system, relying solely on algorithms and protocols for validation. \textit{Proof of Relay} acts like numerous vertical lines, continuously submitting information to the horizontal blockchain formed by \textit{Proof of Availability}, ensuring efficient and reliable message transmission.
\item \textbf{Proof of Availability}: \textit{Proof of Availability} in SendingNetwork plays the role of network-wide consensus, forming the backbone, the horizontal ``chains'' of the network. Through \textit{Proof of Availability}, network validators participate in the generation and verification of blocks by staking tokens, thereby ensuring the overall security and stability of the network. This mechanism not only improves energy efficiency and lowers entry barriers but also promotes positive actions among network participants through incentive and penalty mechanisms, providing a solid infrastructure for SendingNetwork.
\end{itemize}

Through this innovative combination of consensus mechanisms, SendingNetwork can optimize its performance and scalability while maintaining decentralization and network security, ensuring the network can adapt to the evolving future demands and challenges. This comprehensive consensus architecture is a key step in SendingNetwork's journey towards efficient and sustainable blockchain network development.

\section{Concrete Implementation}
\label{sec:cc}
In this section, we demonstrate the detailed protocol construction and framework implementation.
\subsection{Decentralized Identity Protocol}
In the edge network, Decentralized Identifiers (DIDs) serve as a digital identity mechanism to authenticate the identities of individuals engaged in communication \cite{dib2020decentralized}. This begins with the generation of a random key pair when the client starts, followed by the updating of the public key in the DID document once a user signs in using their master or sub-wallet. The master wallet address associated with the DID is then utilized as the user's identifier. 

\begin{figure}[htbp]
    \centering
    \includegraphics[width=0.48\textwidth]{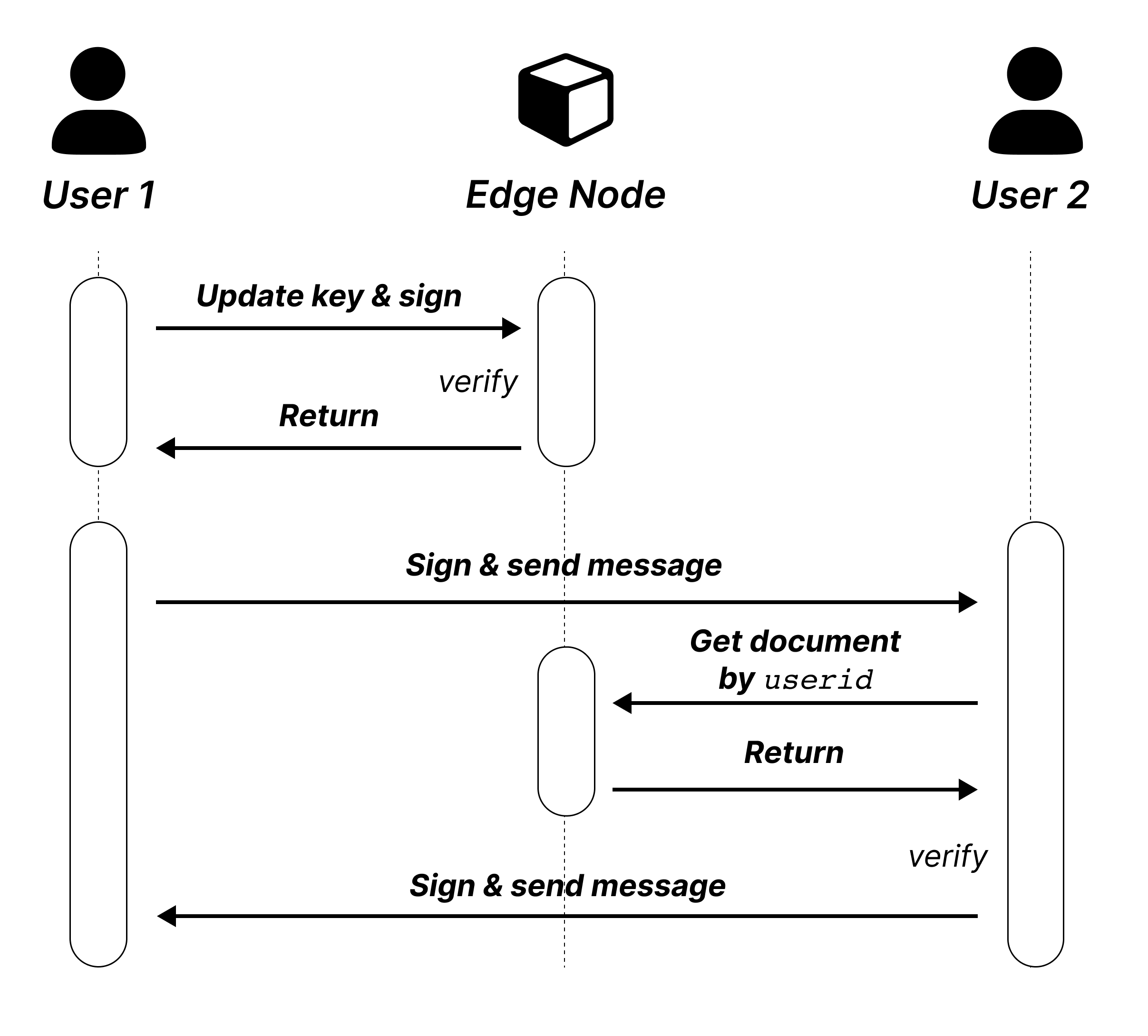}
    \caption{Authentication and Verification}
    \label{fig:authentication_verification}
\end{figure}

For message sending, the user signs the message with the client-generated private key and forwards it to the intended recipient. Upon receipt, the recipient retrieves the sender's DID document from the edge node, identifying it via the sender's master wallet address, which is linked to the user ID. The next step involves the recipient verifying the integrity of this document by inspecting the \textit{controllerSignature} and \textit{keySignature}. If both the document and the signature are authenticated as legal, the recipient can confidently ascertain the sender's identity and accordingly accept the message. This process ensures a secure and verified communication channel within SendingNetwork, leveraging decentralized identity to bolster trust and security among users. The authentication and verification processes are illustrated in Figure~\ref{fig:authentication_verification}.

\subsection{Reliable Messaging}
Rooms are the primary units for conversations within the network. They can facilitate both direct messaging between two users and group chats. They are distributed in nature, with each participant's node(s) maintaining a copy of the room's state.

SendingNetwork functions primarily as a relay network, not retaining persistent message storage for users. It temporarily caches messages for offline users, but these messages are deleted once retrieved. Users have the option to back up their social data, such as message history, contact lists and social network connections, on decentralized platforms like IPFS or using offchain storage solutions. This approach maintains the efficiency and lightness of the network while providing users with data backup options.

A fundamental aspect of the protocol is its commitment to privacy and security, prominently showcased through its implementation of end-to-end message encryption. In scenarios where the group's composition changes – for instance, when a participant leaves the room – the encryption keys are dynamically updated. This update occurs at each participant's local encryption session, ensuring that the integrity and confidentiality of the group chat are maintained even when its status changes.

This approach not only safeguards messages against unauthorized access but also adapts to the dynamic nature of group communications, ensuring consistent security regardless of changes within the group.

\subsubsection{End-to-end Encryption (E2EE)}
End-to-end encrypted chat Protocols, although sometimes overlapping with other types, are explicitly designed to ensure that messages remain readable solely by the intended recipients.


E2EE in large groups is ideally suited for scenarios where only a few members are actively sending messages while the majority are passive readers. Given these considerations, SendingNetwork optimizes the E2EE algorithm for large groups, as the resource consumption for key exchanges becomes impractical. The threshold of such a group is subject to real-time adjustment.

\subsubsection{Encryption Key Exchange}
The protocol employs the \textit{Double Ratchet} algorithm for secure key exchange. Each client device maintains an encryption session locally. When entering a chat group, a key exchange occurs with all other group members except for the client's own device. The key exchange process itself is conducted through asymmetric encryption, ensuring its resistance to decryption attempts.

After the key exchange is completed, each member in the group generates a keychain for every other member, resulting in a total of $(N-1)$ keychains in addition to their own. These keychains are stored locally on each member's device. Whenever a message needs to be sent, the sender selects the current key from their own keychain for encryption. Conversely, when receiving a message, the recipient utilizes the current key from the sender's corresponding keychain for decryption.

In the event of changes to the group's status, such as a member leaving the group, it becomes necessary to update the encryption keys to ensure that the departing member can no longer decrypt the group's messages. To achieve this, the group members engage in another round of key exchange. it is important to note that due to the time complexity of the key exchange process being proportional to $N^2$, for larger groups, an alternative encryption scheme is employed, which we will discuss in more detail later.


\subsubsection{Dynamic Group Encryption Key Exchange}
The symmetric encryption for messages utilizes the common AES-256 algorithm. The transmission of key $\mathtt{A}$ through asymmetric encryption methods can leverage the existing key transmission process of End-to-End Encryption.

Key $\mathtt{A}$ comprises two parts: ID and the key itself. The ID information of key $\mathtt{A}$ is inscribed in the unencrypted section of the message body. Decryption involves retrieving the actual key based on the key ID and then decrypting the message content.
Each individual in every group maintains a currently utilized key $\mathtt{A}$. Following changes in group membership, a reset message is issued, rendering all currently used key $\mathtt{A}$s within the group null.

When a member within the group sends a message and discovers an empty key $\mathtt{A}$, a new key is generated and synchronized with others. For other members sending messages, if a key $\mathtt{A}$ is found locally, it is directly employed. If not, a new key $\mathtt{A}$ is generated in the end-to-end encryption session and synchronized with others.

In scenarios involving large group chats, where the number of participants notably exceeds a high threshold, this encryption method can be implemented. It streamlines the key exchange process, scaling it down from $O(N^2)$ to $O(N)$, and remains within the framework of end-to-end encryption.

\subsubsection{Message Verification}
Any message from the node undergoes signature authentication. All fields in the event sent by the sender (unless specified otherwise) are signed. Upon receiving, the recipient will verify the signature to confirm that the message was not tampered with during transmission, ensuring the authenticity and integrity of the message.

\subsubsection{Blockchain Extensibility}
Leveraging blockchain accounts, SendingNetwork unlocks a spectrum of on-chain functionalities within chat systems. This feature enriches the chat experience with diverse blockchain operations such as trading and socializing, all accessible via the client chat interface. It empowers developers to seamlessly embed capabilities like transfers, airdrops, bill splitting, and other useful features directly into chats.

Developers are equipped to formulate custom input field commands, specifically designed to interact with chosen decentralized applications. These bespoke commands bridge the gap between the chat interface and blockchain technologies, enhancing user engagement. The process is initiated by client-side commands, setting off the interaction sequence. These commands undergo initial interpretation at the client node level. Following this, the client nodes activate the relevant RPC node services, streamlining the blockchain interaction. 

To aid developers, SendingNetwork introduces a plugin extension protocol. This tool is crafted to streamline the development process, making it easier to integrate and manage dApps within the platform.

\subsection{Edge Relay Network}
Edge nodes are identifiable by their public IP addresses and access ports. Any server that aligns with the specified hardware and network bandwidth criteria can be set up as an edge node. 

Their primary responsibilities encompass relaying messages and caching messages for offline clients. The specific auxiliary services an edge node supports can vary based on the server's performance capabilities. By adjusting various configuration parameters, edge nodes can adopt diverse roles, tailoring their functions to meet the network's needs effectively.

\begin{figure*}[htbp]
    \centering
    \includegraphics[width=0.75\textwidth]{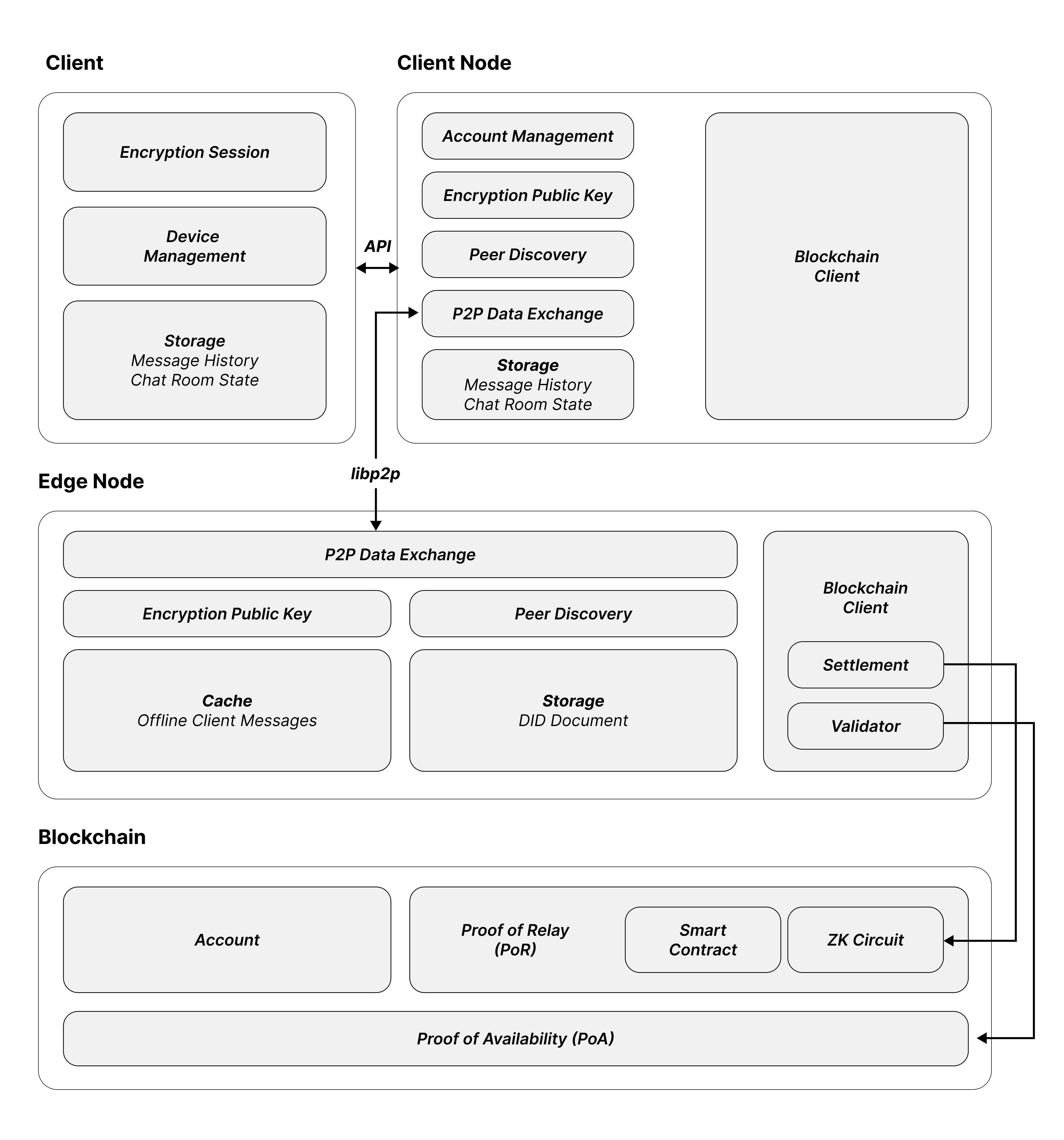}
    \caption{Network Components}
    \label{fig:Network_Components}
\end{figure*}

\subsubsection{Message Relay}
\begin{itemize}
    \item \textbf{Intranet Penetration} \\
In the SendingNetwork framework, the transmission and reception of messages occur over a decentralized P2P network. Situations, where users are within different intranets or in mutually inaccessible networks, necessitate the intervention of servers known as edge nodes. These edge nodes, possessing static addresses, are crucial for relaying messages. SendingNetwork’s P2P protocol inherently supports this relay functionality, enabling users to simply activate relevant settings on their connected edge nodes, thereby designating them as the agents for message forwarding. This setup facilitates the seamless transfer of messages across network barriers. Edge Nodes can also function as WebRTC streaming servers, offering relay services for both voice and video communications.

\item \textbf{Group Chat Optimization} \\
Standard P2P protocols typically transmit messages to peers by relaying them across multiple nodes, often resulting in increased network traffic and latency, and sometimes causing broadcast storms. SendingNetwork addresses these issues by optimizing the network topology specifically for group chat message broadcasting. In this optimized system, group chat messages are primarily processed and conveyed through selected edge nodes. Other edge nodes participate in a supporting capacity to maintain network security and activity, thereby minimizing redundant message flooding and alleviating the load on the network.

Additionally, users or organizations can deploy their own edge nodes, using these exclusively for their messaging needs. For enhanced security, these edge nodes can be situated within a firewall, ensuring safe intra-network communication while maintaining connectivity with the broader network. These specially appointed edge nodes can also be designated to manage group chats. They efficiently forward all group chat messages from their users, significantly boosting the efficacy of group chat communications.

\item \textbf{Notification} \\
In scenarios where a user is offline, such as when a mobile device is locked, the operating system requires a mechanism to alert the user of incoming messages. In a decentralized system without a central server, this notification feature cannot be solely managed by the client involved in the chat. To address this, users are randomly assigned some edge nodes for each chat, including both group and single chats. Notifications are then relayed to the user's device from these edge nodes. For third-party applications integrated with SendingNetwork, the edge node can transmit notification data to the push server of the third-party software provider. The push notification functionality is then executed by the third-party server, ensuring users are promptly informed of new messages.
\end{itemize}

\subsubsection{Room State Management}
SendingNetwork manages room state across various clients using a decentralized approach that ensures consistency and order, even in the absence of a central authority. This is achieved through a series of mechanisms designed to handle the complexities of distributed communication and state management.

The protocol uses a Directed Acyclic Graph (DAG) to maintain the history of all events (messages, state changes) in a room. Each event references one or more previous events, creating a graph structure. This ensures that every participant has a record of the entire history and can trace the sequence of events.

When a user sends a message or alters the room state, this event is propagated to all other participants in the room. Due to network delays or other factors, these events might reach different users at different times.

The room state resolution algorithm manages conflicting states. When different clients end up with different views of the room state due to receiving events in different orders, these algorithms kick in to resolve the conflicts and ensure all clients agree on a single, consistent view of the room state.

The state resolution algorithm employed is deterministic. This means that given the same set of events, every client will calculate the same resulting room state, ensuring consistency across the network. In scenarios where network issues cause the event graph to fork, the state resolution algorithm efficiently merges these forks, ensuring a consistent and unified room state for all participants.

By leveraging these mechanisms, SendingNetwork maintains a coherent and consistent room state across different clients, efficiently handling the challenges posed by distributed, real-time communication. This approach ensures that even in the face of network delays, node outages, or other disruptions, the integrity and continuity of the chat room are preserved.

\subsection{Decentralized Data Storage}
The Edge network does not offer long-term storage capabilities. While a caching mechanism exists to enhance the user messaging experience, the service remains functional even without the cache. All user-related data is primarily stored either on the local device or the Delegation node. This design intentionally minimizes reliance on Edge nodes, considering their third-party nature within a trustless environment.

\subsubsection{User Chat \& Profile Data Storage}
User chat histories and profile data are predominantly stored locally on the user's device. This approach offers several advantages, including data ownership and control. By keeping data on the user's device, SendingNetwork empowers users to maintain possession of their chat conversations and profile information. This data can be encrypted and secured using robust security measures to prevent unauthorized access.

To enhance accessibility and availability, user data can be securely stored on Delegation nodes. Delegation nodes are strategically positioned to ensure redundancy and efficient data retrieval. Users can access their data from any device by connecting to a Delegation node, providing a seamless experience even when transitioning between devices. Data on Delegation nodes is encrypted to maintain user privacy and security.

While the Edge network does not offer long-term storage capabilities, it does temporarily cache the messages for offline clients, which will be deleted upon client retrieval. By adopting this storage strategy, SendingNetwork ensures that user chat and profile data are resiliently stored on the user's device and Delegation nodes, reducing reliance on third-party Edge nodes within the trustless environment.

\subsubsection{Room-related Data Storage}
Essential room-related data, such as room configurations, membership details, and chat histories, are stored locally on the devices of room participants. This decentralized approach ensures that participants maintain control over their room data, promoting data ownership and privacy. Users can access room data even in the absence of a connection to the relay network, ensuring continuous communication.

For improved availability and redundancy of data related to rooms, users who opt to connect to a Delegation node can have a secure copy stored on these nodes. This redundancy ensures that room data remains accessible from various devices and geographical locations. Delegation nodes maintain encrypted copies of room data, preserving user privacy.

\begin{figure*}[htbp]
    \centering
    \includegraphics[width=0.8\textwidth]{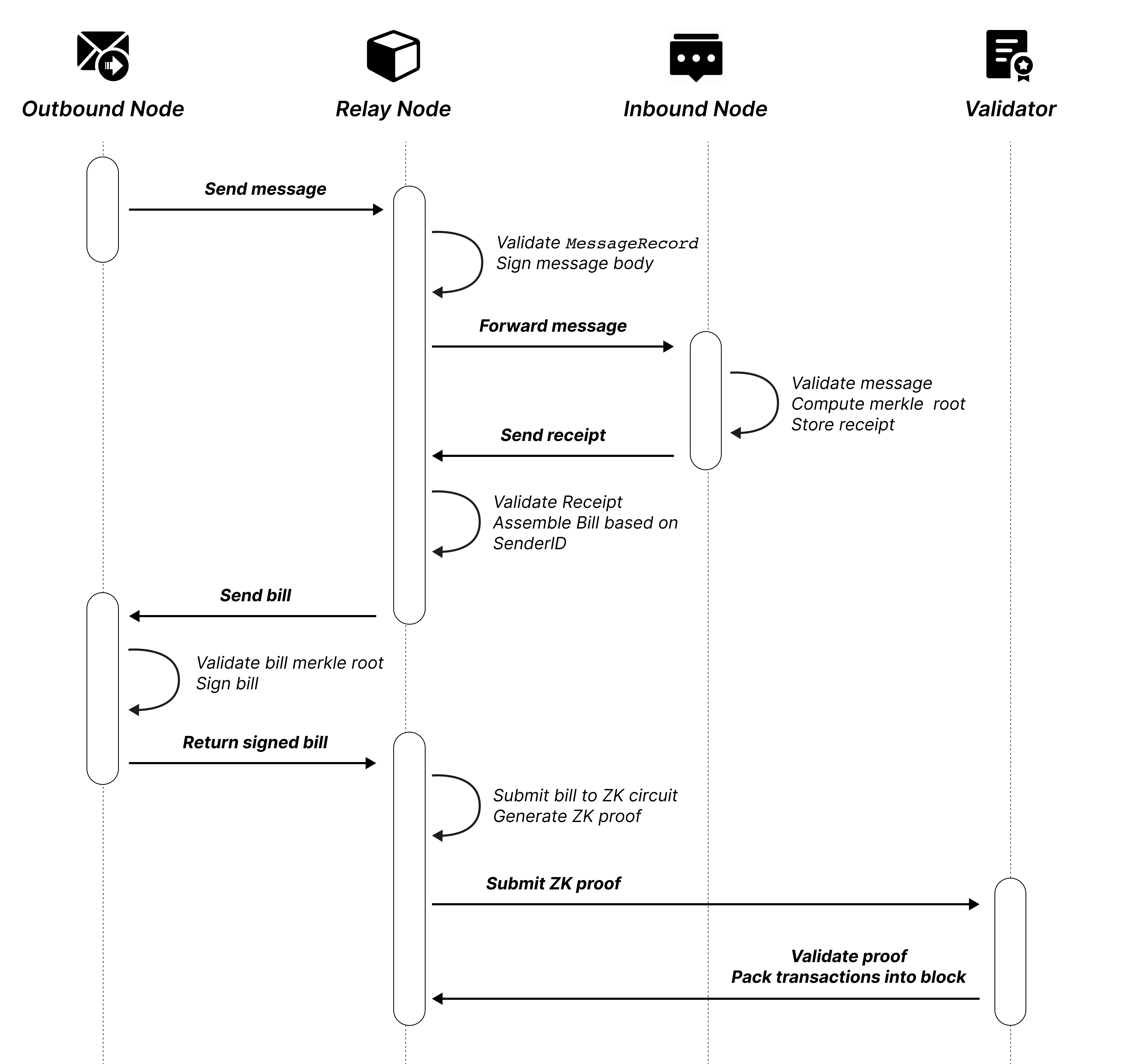}
    \caption{Lifetime of a Message}
    \label{fig:message_process}
\end{figure*}

\section{Designed Consensus Mechanism}
\label{sec:cm}
In this section, we introduce two consensus mechanisms.
\subsection{Proof of Relay}
\textit{Proof of Relay} operates as a localized consensus mechanism, coordinating the monetary interaction of outbound, relay, and inbound nodes in the message transmission process. It validates the successful relay of messages by relay nodes. In this protocol, each outbound node is required to compensate the relay node for its services, and the relay node, in turn, may remunerate any subsequent relay node involved in the process. This financial exchange continues until the message reaches the client node of the intended recipient client, at which point the role of the \textit{Proof of Relay} ceases.

Relay nodes leverage peer discovery protocols to identify the next appropriate node in the relay sequence, facilitating uninterrupted message progression. Consequently, each node in the network may act both as a recipient and a payer in the context of relay fees, serving dual operational roles.
\subsubsection{Elementary Components}
Within this framework, three primary roles are identified:
\begin{itemize}
    \item Outbound Node: The initiator of message relay within a particular segment.
    \item Relay Node: The edge node responsible for the actual relay of the message.
    \item Inbound Node: The receiver of the message from the outbound node in a given relay segment.
\end{itemize}

It is important to note that the \textit{Proof of Relay} mechanism is specifically designed for the interactions between client nodes and relay nodes, and does not pertain to workload verifications between clients and client nodes. This mechanism ensures a reliable relay process. The relay node commits to transmitting the message to designated inbound nodes and then confirms a successful relay to the outbound node. Completion of each relay segment is verified by a signed receipt from the inbound node, thus maintaining the integrity and traceability of the message relay within the network.

Before initiation of the relay process, the message undergoes precise encapsulation. This involves the sender client encrypting the content, followed by the assembly of the message into an event format. The overall processes are explained as follows.
\begin{enumerate}
    \item \textbf{Message Assembly}. The outbound node signs the message, obtaining the signature \textsf{Sig\textsubscript{out}}, and then transmits the message to the relay node. Consequently, the message adheres to the structure:

\begin{equation}
	[EventID + MessageSize]Sig_{out}
\end{equation}

\item \textbf{Message Transmission}. The outbound node transmits the message to the relay node, which then verifies the message's legitimacy. The relay node's verification involves the following steps:

\begin{enumerate}
  \item Validation against \texttt{RoomMemberlist} to confirm appropriate delivery.
  \item Authentication of the message signature to ascertain the match with the outbound node.
  \item Post-verification, the relay node records the message size, endorses the message body with \textsf{Sig\textsubscript{relay}}, extracts the next hop information, and forwards it to the inbound node. 
\end{enumerate}

The message acquires the format:
 \begin{equation}
 	[[EventID + Message Size]Sig_{out}]Sig_{relay}
 \end{equation}

\item \textbf{Receipt Transmission}. Upon receipt, the inbound node undertakes the following validation process:
\begin{enumerate}
  \item Checking \texttt{RoomMemberlist} for delivery appropriateness.
  \item Verifying \textsf{Sig\textsubscript{out}} and \textsf{Sig\textsubscript{relay}} for authenticity.
\item Post-verification, the inbound node decrypts the message, computes the Merkle tree root, stores the receipt, and returns it to the relay node upon reaching a threshold of a certain number of signed messages or after a certain amount of time.
\end{enumerate}
The receipt is endorsed with \textsf{Sig\textsubscript{in}} and structured as:
 \begin{equation}
 \begin{aligned}
	[OutboundNodeID, & EventID, CostCoefficient, \\
 & MerkleRoot]Sig_{in} 
 \end{aligned}
 \end{equation}

\item \textbf{Bill Transmission}. The relay node, upon receipt of the inbound node's receipt, performs legitimacy checks:
\begin{enumerate}
  \item Ensuring receipt of all messages as indicated by \textsf{EventID}.
  \item Verifying \textsf{Sig\textsubscript{in}} to confirm the inbound node's authenticity.
\item Post-verification, the relay node compiles receipts into a bill using \textsf{OutboundNodeID}. The outbound node then examines the bill, focusing on the Merkle root for accuracy assessment and verifying \textsf{Sig\textsubscript{relay}} for authenticity. The outbound node endorses the bill with \textsf{Sig\textsubscript{in}} and forwards it back to the relay node.
\end{enumerate}
The bill is structured as:
 \begin{equation}
 \begin{aligned}
	[EventID, & MessageSize, CostCoefficient, \\
 &MerkleRoot]Sig_{in}
  \end{aligned}
\end{equation}

\item \textbf{ZK Proof Generation}. The relay node generates a zero-knowledge proof via the ZK circuit and submits it to the validator. Upon successful validation, the relay fee is incorporated into a transaction in the latest blockchain block.
\end{enumerate}

\subsubsection{Proof of workload based on KZG commitment}
SendingNetwork builds decentralized relay networks to provide message delivery acceleration services and edge computing services to the system. Relay nodes need to be rewarded for providing bandwidth and arithmetic services, and the system needs a reasonable scheme for workload validation of relay nodes.

KZG Polynomial Commitments is a scheme for committing to polynomials \cite{momose2023security,zhang2022polynomial}. KZG commitments provide functionality similar to Merkle tree roots and can be used to prove that certain data exists in a certain data set, e.g., that a certain transaction is within a certain transaction set.

For single communication relaying, we first consider proving that the relaying node relayed a valid message. The mechanism is composed of the following four Probabilistic Polynomial Time (\textbf{PPT}) algorithms.
\begin{itemize}
    \item $\textbf{Setup}(1^\kappa, t)$ computes two groups $G$ and $G_1$ of prime order $p$ (providing $\kappa$-bit security) such that there exists a symmetric bilinear pairing $e : G \times G \rightarrow G_T$ We denote the generated bilinear pairing group as $G_T = \langle e, G, G_1 \rangle$. Choose a generator $g \in G$, g is the generator on the elliptic curve G. Let $\alpha \in \mathbb{Z}_p^*$ be $SK$, generated by a (possibly distributed) trusted authority. 

\textit{Setup} also generates a $(t + 1)$-tuple $\langle g, g^\alpha, \ldots, g^{\alpha_t} \rangle \in G^{t+1}$ and outputs $PK = \langle G, g, g^\alpha, \ldots, g^{\alpha_t} \rangle$. $SK$ is not required in the rest of the construction.
\item $\textbf{Commit}(PK, \phi(x))$ computes the commitment $C = g^{\phi(\alpha)} \in G$ for polynomial $\phi(x) \in \mathbb{Z}_p[X]$ of degree $t$ or less. For $\phi(x) = \sum_{j=0}^{deg(\phi)} \phi_j x^j$, it outputs $C = \sum_{j=0}^{deg(\phi)} g^{\alpha_j} \phi_j$ as the commitment to $\phi(x)$.
\item  $\textbf{CreateWitness}(PK, \phi(x), i)$ computes $\psi_i(x) = \phi(x) - \phi(i)$ and outputs $\langle i, \phi(i), w_i \rangle$, where the witness $w_i = g^{\psi_i(\alpha)}$ is computed in a manner similar to $C$, above.
\item $\textbf{VerifyEval}(PK, C, i, \phi(i), w_i)$ verifies that $\phi(i)$ is the evaluation at the index $i$ of the polynomial committed to by $C$. If $e(C, g) == e(w_i, g^\alpha/g^i)e(g, g)^{\phi(i)}$, the algorithm outputs $1$, else it outputs $0$.
\end{itemize}

The overall process mainly uses the KZG commitment to verify that the relay node has indeed signed message i. However, this process is completely unnecessary because the process only serves to protect the privacy of the relay node. However, this process is completely unnecessary because the KZG only serves to protect the privacy of the relay node, and in fact the process of relaying over a message by a relay node does not require privacy protection. So the workload proof for single communication relaying is meaningless.

We consider the simultaneous verification of n message relay processes, and the contents of the n message relay processes form the vector $v = [v_0, v_1, \dots, v_{n-1}]$, one can set the interpolating polynomial $\phi(X)$ such that $\phi(i)=v_i$ as follows:

 \begin{equation}
 \begin{aligned}
    & \phi(X) = \sum_{i=0}^{n-1} v_i \cdot \psi_i(X), \\
    \\
    & \text{where}\; \psi_i(X) = \prod_{\substack{j\in [0,n)\\\\j\ne i}}\left(\frac{X-j}{i-j} \right)
\end{aligned}
\end{equation}
$i$ corresponds to the id of message $v_i$, then the task that KZG promises is that $v_i$ is the $i$th element in the vector.
\begin{itemize}
\item $\textbf{Setup}(1^\kappa, t)$ still takes an elliptic curve pairing at prime order p. The pairing of $\tau$ is a random number. $\tau$ is an arbitrary random number, and the generated public parameter $PK=\left(g^{\tau^i},\ell_i\right)_{i\in[0,n)}$, where $l_i = g^{\psi_i(\tau)}$, and $(g,g^{\tau})$ in the public parameter serves as a validation key for the proof of commitment.

\item $\textbf{Commit}(PK, \phi(x))$ computes the commitment $C=g^{\phi(\tau)}$, which is computed as
 \begin{equation}
 \begin{aligned}
   C &= \sum_{i=0}^{n-1} l_i^{v_i} = \sum_{i=0}^{n-1} g^{v_i \cdot \psi_i(\tau)}\\
   &= g^{\prod_{i=0}^{n-1} v_i \cdot \psi_i(\tau)} \\
   &= g^{\phi(\tau)}
\end{aligned}
\end{equation}

\item $\textbf{CreateWitness}(PK, \phi(x), \alpha)$ We take a challenge point $(\alpha, \phi(\alpha))$ to generate the witness. First calculate the quotient $q_\alpha(\tau) = \frac{\phi(\tau) - \phi(\alpha)}{\tau-\alpha}$ , the witness will be $\pi_\alpha=g^{q_i(\tau)} = g^ \frac{\phi(\alpha)}{\tau-\alpha}$ , witness will be $\pi_i=g^{q_\alpha(\ tau)} = g^ \frac{\phi(\tau)-\phi(\alpha)}{\tau-\alpha}$.

\item $\textbf{VerifyEval}(PK, C, \alpha, \phi(\alpha), \pi_\alpha)$ verifies that
 \begin{equation}
 \begin{aligned}
   e(C/g^{v_\alpha}, g)=e(\pi_\alpha, g^{\tau}/g^\alpha)
\end{aligned}
\end{equation}
At this point the four main functions have actually accomplished job verification for the relay nodes. The Aggregatable Subvector Commitments provide additional functionality, including commitment updates and I-subvector proofs.

\item $\textbf{Commitment renewal:}$ KZG commitments and thus vector commitments are homomorphic: given commitments $C$ and $C'$ to ${v}$ and ${v'}$, we can get a commitment $C=C \cdot C'$ to ${v} + {v'}$. A consequence of this is that we can easily update a commitment $c$ to $C'$, given a change $\delta$ to $v_i$ as:
 \begin{equation}
 \begin{aligned}
C' = C \cdot l_i^{\delta}
\end{aligned}
\end{equation}
This feature makes our commitment process more continuous and reliable. At the beginning of each validation epoch, initialize the vector $v = \{0\}_n$, and each completed message relay is equivalent to $v \gets v + \vec v_i$, which corresponds to a change in the commitment as 
 \begin{equation}
 \begin{aligned}
C \gets C * g^ {v_i \cdot \psi_i(\tau)}
\end{aligned}
\end{equation}

\item $\textbf{Subvector batch proofs:}$ To increase the credibility, multiple sampling proofs are made for the commitment of a relay node, let the sampling vector $(v_i)_{i\in I}$, an $I$-subvector proof $\pi_I$ can be computed using a KZG batch proof as:
 \begin{equation}
 \begin{aligned}
\pi_I &= g^{q_I(\tau)}=g^\frac{\phi(\tau)-R_I(\tau)}{A_I(\tau)}
\end{aligned}
\end{equation}
Verifying the proof can also be done with two pairings:
 \begin{equation}
 \begin{aligned}
e(c/g^{R_I(\tau)}, g)=e(\pi_I, g^{A_I(\tau)})
\end{aligned}
\end{equation}
where I is the set of sampling points, and
 \begin{equation}
A_I(X) =\prod_{i\in I} (X - i)
\end{equation}

\begin{equation}
R_I(X) =\sum_{i\in I} v_i \left(\prod_{j\in I,j\ne i}\frac{X - j}{i - j}\right)
\end{equation}

\end{itemize}
In the above process, we use the KZG commitment to verify whether the relay node succeeded in relaying the message or not, in practical application scenarios, we will have two additional options. One is to use the Plonk algorithm to circuitize the validation process and accelerate the generation and validation of the promises. The other option is to apply zkEVM Validium directly, which is currently available on Polygon. We favor a fully CIRCUIT implementation.

\subsection{Proof of Availability}
\textit{Proof of Availability} is a consensus mechanism that allows edge nodes to participate in the block generation process by engaging in staking activities.

\subsubsection{Validator Selection}
Within \textit{Proof of Availability}, validator selection is governed by the \textit{RANDAO} pseudorandom function \cite{alturki2020statistical}, applied to staking edge nodes. This Proof of Stake-based mechanism evaluates both the quantity of staked tokens and the node's operational metrics such as uptime, latency, and overall contribution, ensuring robust network performance and security.

\subsubsection{Transaction Verification}
Elected validators, chosen from edge nodes, are tasked with processing transaction requests. These include token transfers, smart contract executions, and other network functions. Validators are responsible for confirming the legality of transactions, which entails signature verification, double-spending prevention, and smart contract rule adherence.

\subsubsection{Proof of Relay Assessment}
For network security and stability, validators are required to assess the submission of \textit{Proof of Relay} by edge nodes in recent blocks. This evaluation is pivotal in determining an edge node's active participation in message relay, a critical measure of service quality.

In the POR design, we use KZG to complete the workload verification of whether the relay node has relayed the message or not, but there is still an issue to be dealt with at the consensus level. The message delivery process will go through multiple nodes relaying, this process may have malicious nodes stealing previous relay node's work and sending fake verification to steal rewards, so we designed the chain signature system. A signature field is added within the message passing STRUCTURE which includes the signature result of each relay. The signing process is that the current relay node uses the private key to sign the message ID, the signature of the previous hop node, and the public key of the next hop node.
\begin{figure}[h]
  \begin{minipage}{1\linewidth}
    \centering
    \includegraphics[width=0.6\textwidth]{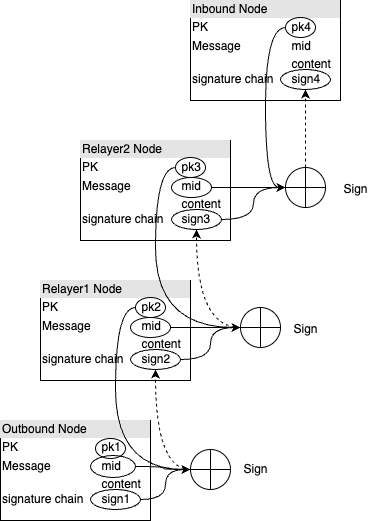}
    \caption{Chained Signatures}
    \label{fig:chain_sign}
  \end{minipage}
\end{figure}

\subsubsection{Verkle Trees}
Verkle trees (a portmanteau of "Vector commitment" and "Merkle Trees") are a data structure that can be used to upgrade Ethereum nodes so that they can stop storing large amounts of state data without losing the ability to validate blocks.

Since we use KZG to generate vector promises and proofs in the main proof-of-work protocol of \textit{Proof of Relay}, it makes our system naturally suitable for using the Verkle Tree structure. We plan to use KZG to generate verifications for all transactions except for relay verifications, and then form the blockchain's storage structure with promises and proofs.

\begin{figure}[h]
  \begin{minipage}{1\linewidth}
    \centering
       \includegraphics[width=0.9\textwidth]{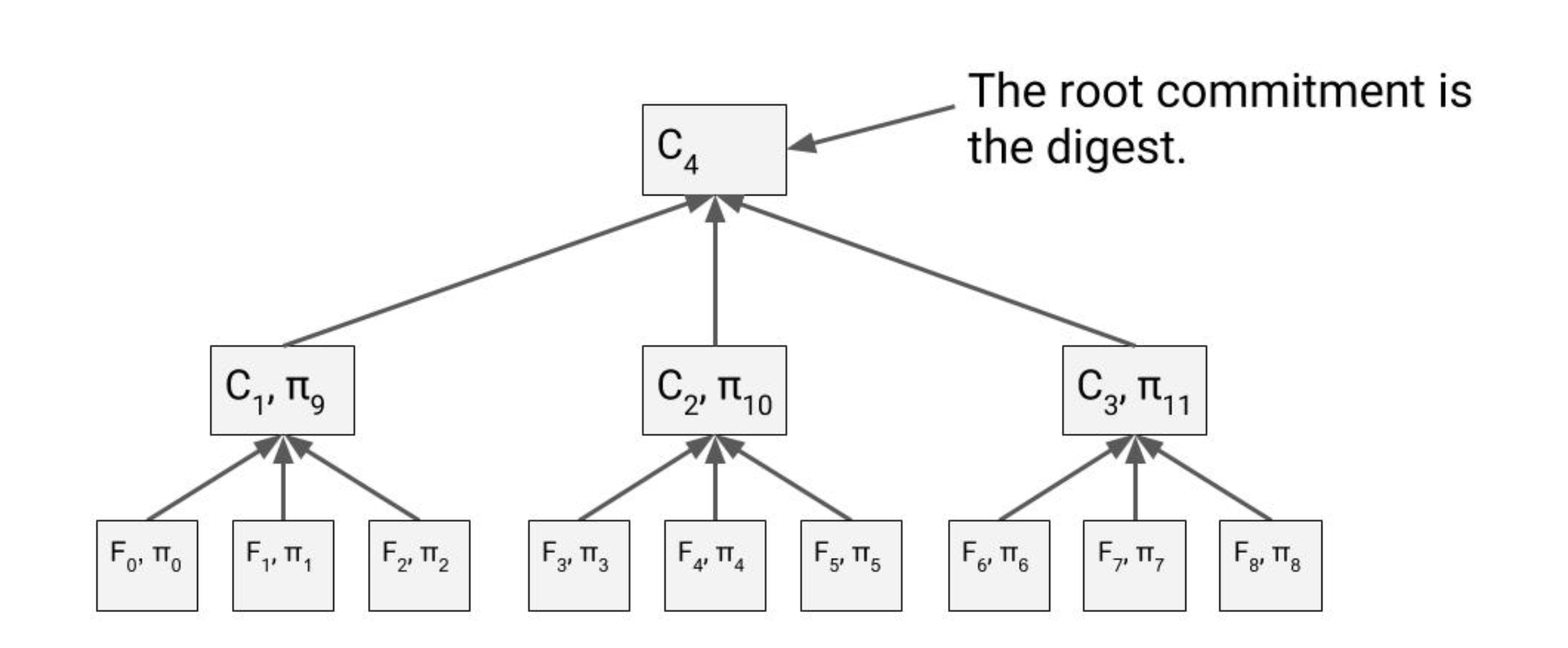}
    \caption{Verkle Trees}
    \label{fig:tree}
  \end{minipage}
\end{figure}

\section{Future Work}
\label{sec:fw}
In this section, we discuss the scheduled technical route and upcoming novel features slated for future integration.
\subsection{Technical Roadmap}
Utilizing autonomous intercommunication and networking amongst Edge Nodes, a vast Decentralized Physical Infrastructure (DePIN) for communication emerges. This infrastructure's integrity and sustainability are assured through the application of encryption, Delegation, and Edge-based relay techniques, ensuring the security of the network formed by various devices. Concurrently, the economic layer of network development on the blockchain is fortified, focusing on authenticating the veracity of Edge Nodes in the physical world and validating the reliability of relay data. Our vision and its realization involve a significant amount of work to achieve. Currently, the roadmap to achieve this vision is divided into five stages:

\begin{enumerate}
  \item \textbf{Foundational Data and Communication:} Enable communication capabilities that coexist with P2P, Delegation, and Private modes based on a P2P network. Construct an edge relay network and implement the room state synchronization and message caching mechanisms.
  \item \textbf{End-to-End Encryption:} Implement complete \textit{Double Ratchet}-based end-to-end encryption and signature verification for communication data to ensure data security and privacy.
  \item \textbf{Proof of Relay Consensus:} Employ smart contracts on a third-party chain along with challenge nodes to track miners' workload.
  \item \textbf{Proof of Availability Consensus:} Build a global consensus blockchain based on Proof of Stake, migrating all workload tracking to the newly constructed blockchain.
  \item \textbf{Mainnet Launch:} Following comprehensive product testing, launch the main network.
\end{enumerate}

Currently, we have made preliminary progress in the first two stages. Moving forward, alongside refining communication performance and security, we will commence work related to consensus and tokenomics design.

\subsection{Other Features}
Apart from the consensus-related works, there are other features we need to take into consideration. The goal is to create scalability for the network and provide better service to a broader community.

\subsubsection{Multichain Support}
SendingNetwork recognizes the importance of interoperability across blockchain networks \cite{umran2023multi}. To achieve this, we are actively working on incorporating multichain support \cite{kovst2020multi,abuidris2022collaborative}. This milestone will enable users to seamlessly interact with SendingNetwork across various blockchain platforms, increasing accessibility and versatility.

\subsubsection{Burn-after-Reading}
"Burn after reading" offers enhanced privacy and security by ensuring the automatic deletion of messages after being read, addressing the growing concern for data privacy in digital communication \cite{xiong2021burn,yang2022burn,zou2020burn}. This feature is significant for its role in safeguarding sensitive information, reducing the risks associated with data breaches, and promoting open communication, especially in scenarios involving confidential data exchange. It aligns with legal data protection requirements, aiding businesses in regulatory compliance. However, implementing this concept poses technical challenges, including ensuring reliable and verifiable message deletion across diverse devices and platforms, managing unread messages, and preventing the exploitation of malicious activities. Additionally, it involves balancing user experience with privacy needs and navigating complex legal and ethical implications. We will leverage some advanced techniques, such as puncturable encryption \cite{susilo2020puncturable,cui2023secure} and proxy-based re-encryption \cite{maiti2020p2b} to construct a novel cryptographic primitive that achieves the fantastic ``Burning after reading'' feature for our users in the future.

\subsubsection{Account Abstraction for Smart Contracts}
As part of our commitment to offering a comprehensive communication and utility platform, SendingNetwork will introduce Account Abstraction (AA) \cite{singh2023account,huang2022ethereum} support. AA will empower users to interact with smart contracts directly through SendingNetwork, bridging the gap between messaging and blockchain applications. AA's assistance enables the development of functionalities such as Social Recovery, a useful tool for account recovery purposes.

Our vision extends beyond communication to encompass broader financial and economic interactions. SendingNetwork's roadmap includes the creation of decentralized spending accounts, enabling users to manage their digital assets and engage in financial activities securely and autonomously within our ecosystem.

\section{Conclusion}
\label{sec:cl}
In the current era, while the computing and storage domains witness innovation, the communication sector faces numerous challenges. Shortcomings in social communication protocols and technologies highlight users' vulnerabilities in the digital realm. These concerns span data ownership, account ownership, privacy and security, transaction safety, as well as platform fragmentation and closure. Addressing these challenges requires an urgent decentralized communication protocol to ensure users possess ownership over their data and accounts, fostering an open social ecosystem.

In response, the inception of SendingNetwork aims to fulfill these needs by constructing an incessant, self-sustaining, private, and scalable communication protocol. Employing a three-tiered architecture of access-relay-consensus, coupled with mechanisms like end-to-end encryption, Delegation, \textit{Proof of Relay}, and \textit{Proof of Availability}, SendingNetwork exhibits characteristics of decentralization, privacy, security, and high efficiency. This architecture leverages the strengths of current decentralized communication technologies while mitigating their shortcomings, empowering users with greater ownership and control.

Overall, SendingNetwork not only transcends the technical limitations of traditional communication protocols but also establishes a genuinely usable, decentralized, and securely private digital social platform. In our future roadmap, we aim to continually refine our technical solutions, gradually achieving goals like consensus mechanisms and mainnet launch, ensuring users experience a safer and more efficient communication environment.


%


\ifCLASSOPTIONcaptionsoff
  \newpage
\fi



%

\bibliographystyle{IEEEtran}
\bibliography{sendingNetwork-whitepaper}

\begin{thebibliography}{10}
\providecommand{\url}[1]{#1}
\csname url@samestyle\endcsname
\providecommand{\newblock}{\relax}
\providecommand{\bibinfo}[2]{#2}
\providecommand{\BIBentrySTDinterwordspacing}{\spaceskip=0pt\relax}
\providecommand{\BIBentryALTinterwordstretchfactor}{4}
\providecommand{\BIBentryALTinterwordspacing}{\spaceskip=\fontdimen2\font plus
\BIBentryALTinterwordstretchfactor\fontdimen3\font minus
  \fontdimen4\font\relax}
\providecommand{\BIBforeignlanguage}[2]{{%
\expandafter\ifx\csname l@#1\endcsname\relax
\typeout{** WARNING: IEEEtran.bst: No hyphenation pattern has been}%
\typeout{** loaded for the language `#1'. Using the pattern for}%
\typeout{** the default language instead.}%
\else
\language=\csname l@#1\endcsname
\fi
#2}}
\providecommand{\BIBdecl}{\relax}
\BIBdecl

\bibitem{murray2023promise}
A.~Murray, D.~Kim, and J.~Combs, ``The promise of a decentralized internet:
  What is web3 and how can firms prepare?'' \emph{Business Horizons}, vol.~66,
  no.~2, pp. 191--202, 2023.

\bibitem{hildenbrandt2018kevm}
E.~Hildenbrandt, M.~Saxena, N.~Rodrigues, X.~Zhu, P.~Daian, D.~Guth, B.~Moore,
  D.~Park, Y.~Zhang, A.~Stefanescu \emph{et~al.}, ``Kevm: A complete formal
  semantics of the ethereum virtual machine,'' in \emph{2018 IEEE 31st Computer
  Security Foundations Symposium (CSF)}.\hskip 1em plus 0.5em minus 0.4em\relax
  IEEE, 2018, pp. 204--217.

\bibitem{williams2019arweave}
S.~Williams, V.~Diordiiev, L.~Berman, and I.~Uemlianin, ``Arweave: A protocol
  for economically sustainable information permanence,'' \emph{arweave. org,
  Tech. Rep}, 2019.

\bibitem{psaras2020interplanetary}
Y.~Psaras and D.~Dias, ``The interplanetary file system and the filecoin
  network,'' in \emph{2020 50th Annual IEEE-IFIP International Conference on
  Dependable Systems and Networks-Supplemental Volume (DSN-S)}.\hskip 1em plus
  0.5em minus 0.4em\relax IEEE, 2020, pp. 80--80.

\bibitem{jhaver2021evaluating}
S.~Jhaver, C.~Boylston, D.~Yang, and A.~Bruckman, ``Evaluating the
  effectiveness of deplatforming as a moderation strategy on twitter,''
  \emph{Proceedings of the ACM on Human-Computer Interaction}, vol.~5, no.
  CSCW2, pp. 1--30, 2021.

\bibitem{yang2020zero}
X.~Yang and W.~Li, ``A zero-knowledge-proof-based digital identity management
  scheme in blockchain,'' \emph{Computers \& Security}, vol.~99, p. 102050,
  2020.

\bibitem{isaak2018user}
J.~Isaak and M.~J. Hanna, ``User data privacy: Facebook, cambridge analytica,
  and privacy protection,'' \emph{Computer}, vol.~51, no.~8, pp. 56--59, 2018.

\bibitem{russell2013mining}
M.~A. Russell, \emph{Mining the social web: data mining Facebook, Twitter,
  LinkedIn, Google+, GitHub, and more}.\hskip 1em plus 0.5em minus 0.4em\relax
  " O'Reilly Media, Inc.", 2013.

\bibitem{shihab2009use}
E.~Shihab, Z.~M. Jiang, and A.~E. Hassan, ``On the use of internet relay chat
  (irc) meetings by developers of the gnome gtk+ project,'' in \emph{2009 6th
  IEEE International Working Conference on Mining Software Repositories}.\hskip
  1em plus 0.5em minus 0.4em\relax IEEE, 2009, pp. 107--110.

\bibitem{scalability_p2p_systems}
E.~De~Souza~e Silva, R.~Le{\~a}o, D.~Menasch{\'e}, and D.~Towsley,
  ``Scalability issues in p2p systems,'' 2014.

\bibitem{jacob2019glimpse}
F.~Jacob, J.~Grash{\"o}fer, and H.~Hartenstein, ``A glimpse of the matrix:
  Scalability issues of a new message-oriented data synchronization
  middleware,'' in \emph{Proceedings of the 20th International Middleware
  Conference Demos and Posters}, 2019, pp. 5--6.

\bibitem{schuster2014global}
D.~Schuster, P.~Grubitzsch, D.~Renzel, I.~Koren, R.~Klauck, and M.~Kirsche,
  ``Global-scale federated access to smart objects using xmpp,'' in \emph{2014
  IEEE International Conference on Internet of Things (iThings), and IEEE Green
  Computing and Communications (GreenCom) and IEEE Cyber, Physical and Social
  Computing (CPSCom)}.\hskip 1em plus 0.5em minus 0.4em\relax IEEE, 2014, pp.
  185--192.

\bibitem{peer_to_peer_social_networks}
N.~Masinde and K.~Graffi, ``Peer-to-peer-based social networks: A comprehensive
  survey,'' \emph{SN COMPUT. SCI.}, vol.~1, p. 299, 2020.

\bibitem{liu2020blockchain}
Y.~Liu, D.~He, M.~S. Obaidat, N.~Kumar, M.~K. Khan, and K.-K.~R. Choo,
  ``Blockchain-based identity management systems: A review,'' \emph{Journal of
  network and computer applications}, vol. 166, p. 102731, 2020.

\bibitem{arndt2019decentralized}
N.~Arndt and M.~Martin, ``Decentralized collaborative knowledge management
  using git,'' in \emph{Companion Proceedings of The 2019 World Wide Web
  Conference}, 2019, pp. 952--953.

\bibitem{meffert2009bilinear}
D.~Meffert, ``Bilinear pairings in cryptography,'' \emph{Computing Science
  Department, Radboud Universiteit Nijmegen}, vol. 6525, 2009.

\bibitem{galbraith2016recent}
S.~D. Galbraith and P.~Gaudry, ``Recent progress on the elliptic curve discrete
  logarithm problem,'' \emph{Designs, Codes and Cryptography}, vol.~78, pp.
  51--72, 2016.

\bibitem{perrin2016double}
T.~Perrin and M.~Marlinspike, ``The double ratchet algorithm,'' \emph{GitHub
  wiki}, p. 112, 2016.

\bibitem{alwen2019double}
J.~Alwen, S.~Coretti, and Y.~Dodis, ``The double ratchet: security notions,
  proofs, and modularization for the signal protocol,'' in \emph{Annual
  International Conference on the Theory and Applications of Cryptographic
  Techniques}.\hskip 1em plus 0.5em minus 0.4em\relax Springer, 2019, pp.
  129--158.

\bibitem{guidi2021libp2p}
B.~Guidi, A.~Michienzi, and L.~Ricci, ``A libp2p implementation of the bitcoin
  block exchange protocol,'' in \emph{Proceedings of the 2nd International
  Workshop on Distributed Infrastructure for Common Good}, 2021, pp. 1--4.

\bibitem{dib2020decentralized}
O.~Dib and K.~Toumi, ``Decentralized identity systems: Architecture,
  challenges, solutions and future directions,'' \emph{Annals of Emerging
  Technologies in Computing (AETiC), Print ISSN}, pp. 2516--0281, 2020.

\bibitem{momose2023security}
A.~Momose, S.~Das, and L.~Ren, ``On the security of kzg commitment for vss,''
  in \emph{Proceedings of the 2023 ACM SIGSAC Conference on Computer and
  Communications Security}, 2023, pp. 2561--2575.

\bibitem{zhang2022polynomial}
J.~Zhang, T.~Xie, T.~Hoang, E.~Shi, and Y.~Zhang, ``Polynomial commitment with
  a $\{$One-to-Many$\}$ prover and applications,'' in \emph{31st USENIX
  Security Symposium (USENIX Security 22)}, 2022, pp. 2965--2982.

\bibitem{alturki2020statistical}
M.~A. Alturki and G.~Ro{\c{s}}u, ``Statistical model checking of randao’s
  resilience to pre-computed reveal strategies,'' in \emph{Formal Methods. FM
  2019 International Workshops: Porto, Portugal, October 7--11, 2019, Revised
  Selected Papers, Part I 3}.\hskip 1em plus 0.5em minus 0.4em\relax Springer,
  2020, pp. 337--349.

\bibitem{umran2023multi}
S.~M. Umran, S.~Lu, Z.~A. Abduljabbar, and V.~O. Nyangaresi, ``Multi-chain
  blockchain based secure data-sharing framework for industrial iots smart
  devices in petroleum industry,'' \emph{Internet of Things}, vol.~24, p.
  100969, 2023.

\bibitem{kovst2020multi}
K.~Ko{\v{s}}t'{\'a}l, ``Multi-chain architecture for blockchain networks,''
  \emph{Information Sciences and Technologies}, vol.~12, no.~2, pp. 8--14,
  2020.

\bibitem{abuidris2022collaborative}
Y.~Abuidris, C.~Wang, and W.~Yang, ``Collaborative multi-chain architecture for
  data transmission across homogeneous blockchain,'' in \emph{2022
  International Conference on Innovations and Development of Information
  Technologies and Robotics (IDITR)}.\hskip 1em plus 0.5em minus 0.4em\relax
  IEEE, 2022, pp. 105--110.

\bibitem{xiong2021burn}
H.~Xiong, L.~Wang, Z.~Zhou, Z.~Zhao, X.~Huang, and S.~Kumari, ``Burn after
  reading: Adaptively secure puncturable identity-based proxy re-encryption
  scheme for securing group message,'' \emph{IEEE Internet of Things Journal},
  vol.~9, no.~13, pp. 11\,248--11\,260, 2021.

\bibitem{yang2022burn}
L.~Yang, M.~Gao, Z.~Chen, R.~Xu, A.~Shrivastava, and C.~Ramaiah, ``Burn after
  reading: Online adaptation for cross-domain streaming data,'' in
  \emph{European Conference on Computer Vision}.\hskip 1em plus 0.5em minus
  0.4em\relax Springer, 2022, pp. 404--422.

\bibitem{zou2020burn}
C.~Zou and J.~Xue, ``Burn after reading: A shadow stack with microsecond-level
  runtime rerandomization for protecting return addresses,'' in
  \emph{Proceedings of the ACM/IEEE 42nd International Conference on Software
  Engineering}, 2020, pp. 258--270.

\bibitem{susilo2020puncturable}
W.~Susilo, D.~H. Duong, H.~Q. Le, and J.~Pieprzyk, ``Puncturable encryption: a
  generic construction from delegatable fully key-homomorphic encryption,'' in
  \emph{Computer Security--ESORICS 2020: 25th European Symposium on Research in
  Computer Security, ESORICS 2020, Guildford, UK, September 14--18, 2020,
  Proceedings, Part II 25}.\hskip 1em plus 0.5em minus 0.4em\relax Springer,
  2020, pp. 107--127.

\bibitem{cui2023secure}
H.~Cui and X.~Yi, ``Secure internet of things in cloud computing via
  puncturable attribute-based encryption with user revocation,'' \emph{IEEE
  Internet of Things Journal}, 2023.

\bibitem{maiti2020p2b}
S.~Maiti and S.~Misra, ``P2b: Privacy preserving identity-based broadcast proxy
  re-encryption,'' \emph{IEEE Transactions on Vehicular Technology}, vol.~69,
  no.~5, pp. 5610--5617, 2020.

\bibitem{singh2023account}
A.~K. Singh, I.~U. Hassan, G.~Kaur, S.~Kumar \emph{et~al.}, ``Account
  abstraction via singleton entrypoint contract and verifying paymaster,'' in
  \emph{2023 2nd International Conference on Edge Computing and Applications
  (ICECAA)}.\hskip 1em plus 0.5em minus 0.4em\relax IEEE, 2023, pp. 1598--1605.

\bibitem{huang2022ethereum}
T.~Huang, D.~Lin, and J.~Wu, ``Ethereum account classification based on graph
  convolutional network,'' \emph{IEEE Transactions on Circuits and Systems II:
  Express Briefs}, vol.~69, no.~5, pp. 2528--2532, 2022.

\end{thebibliography}

%








\end{document}